\documentclass[epj-spec]{svjour}
\usepackage{graphics}
\usepackage{epsfig}
\usepackage{amsmath,mathptm}
\begin{document}
\title{Semiclassical and quantum polarons in crystalline acetanilide}
\author{Peter Hamm\inst{1}\fnmsep\thanks{\email{phamm@pci.unizh.ch}} \and
G. P. Tsironis\inst{2,}\inst{3}\fnmsep\thanks{\email{gts@physics.uoc.gr}} }
\institute{Physikalisch-Chemishes Institut, Universit\"at
Z\"{u}rich, Winterhurerstr. 190, CH-8057, Z\"{u}rich, Switzerland
\and Department d' Estructura i Constituents de la Mat\'eria,
Facultat de F\'{\i}sica, Universitat de Barcelona, Diagonal
647, E-08028 Barcelona, Spain
 \and
Department of Physics, University of Crete and Institute of
Electronic Structure and Laser, FORTH, P.O. Box 2208, Heraklion
71003, Crete, Greece. }

\abstract{ Crystalline acetanilide is a an organic solid with
peptide bond structure similar  to that of proteins.  Two states
appear in the amide I spectral region having drastically different
properties: one is strongly temperature dependent  and disappears
at high temperatures while the other is stable at all
temperatures.  Experimental and theoretical work over the past
twenty five years has assigned the former to a selftrapped state
while the latter to an extended free exciton state.
In this article we review
the experimental and theoretical developments on
acetanilide paying particular attention to issues that are still
pending.  Although the interpretation of the states is
experimentally sound, we find that specific theoretical
comprehension is still lacking.  Among the issues that
that appear not well understood
is  the effective dimensionality of the
selftrapped polaron and free exciton states.
} 
\maketitle
\section{Introduction}
\label{s1}
The first numerical experiment performed by Fermi,
Pasta and Ulam in 1955 proved to be the beginning of an exciting,
non-reductionist  branch of modern science focusing on
nonlinearity and complexity in a variety of physical systems~\cite{note1}.
In this path, the discovery of the soliton in the sixties by Zabusky
and Kruskal and the understanding of many of its mathematical
properties led in the seventies in the exploration for physical
applications. Proposals for soliton modes in Josephson junctions,
optical fibers and other physical systems where suggested at this
time, while
Davydov introduced the novel idea that solitons may have a direct
impact to biology as well~\cite{dav73,dav77,dav79}. More
specifically Davydov proposed that the energy released in the
process of ATP hydrolysis becomes vibrationally selftrapped and
forms a soliton.
The main
feature that made solitons appealing to biology is their property
of dispersiveless transport; if a soliton-like packet of energy
forms in a macromolecule, it may propagate without losses and
enable long range coherent energy transfer.   This feature
could be critical in the internal energetics of proteins where
energy deposited on given sites produces large effects at
relatively distant locations.  Although soliton
propagation experiments in
proteins  were not possible at this time, Careri and
coworkers were investigating  independently crystalline
acetanilide (ACN), a molecular solid
that has peptide bonds and a structure similar
to that of a protein~\cite{car73} (Fig 1).  They found in infrared
absorption experiments that an "anomaly" appears in the amide-I
spectral region manifested through a strongly temperature
dependent spectral peak~\cite{car83,car84} (Fig. 2).  In the early
eighties Scott and collaborators surmised that this peak was
related to a Davydov soliton~\cite{car83,eil84,sco92}.  Davydov's
original soliton idea coupled to the experimental findings of
Careri and theoretical picture of Scott for ACN  lead to a furry
of activity during the rest of the decade as well as in the early
nineties. Brown et al. examined critically the
general  "Davydov soliton" hypothesis
 ~\cite{brown86b,brown86c} while in acetanilide Scott as well
Alexander and
Krumhansl soon realized that it was a small polaron rather than an extended
soliton responsible for the anomalous peak~\cite{ale85,ale86,sco89,sco92}.
The small polaron picture corroborated the neutron and absorption
experiments of Barthes and collaborators.
~\cite{bar89,bar91}.
During the same period, explicit lattice discreetness was introduced in
nonlinear approaches
and, as a result,
a new entity was introduced, viz. that of an intrinsic localized mode
or discrete breather~\cite{sie88}.  The latter modes where seen in many
instances to describe better nonlinear
localization than continuous solitons.
The over thirty-year long history of the
fertile Davydov soliton idea is landmarked presently by a new
generation of experiments; the pump-probe experiments of Hamm and
collaborators have  not only corroborated the acetanilide picture
but also  produced the first experimental sign for nonlinearly
localized states in proteins~\cite{edl02a,edl02b,edl03,edl04}.  In
the present work we will attempt to summarize the basic
theoretical and experimental steps in the Davydov soliton idea as
related specifically to acetanilide.
We prefer to focus only on the
latter since the excellent review by Scott in 1992 covers much of the
generalities on the theoretical Davydov soliton idea,
while,additionally, acetanilide provides
a tangible system where both theory and experiments have been employed.

Nonlinear excitations in biomolecules are typically associated with
with objects such as solitons, polarons or
discrete breathers (DBs). All three modes are similar in nature
in that they arise due to the actual or effective presence of some
type of nonlinearity in the equations of motion.
In the original formulation of the
Davydov problem, a vibrational excitation (a C=O stretching
quantum, or a $C=O$ exciton or a vibron) with typical energy of about
$1665~\rm{cm}^{-1}$  is coupled to lattice phonons with energies
that are more than one order of magnitude smaller.   Due to the
assumed strong exciton-phonon coupling, the bear exciton becomes
self-trapped and forms a polaron, i.e. a new entity
that is localized and has
lower overall energy than the extended exciton.  Depending on the
relative values of the exciton hopping term, the phonon frequency
and the exciton-phonon coupling one may arrive to a polaron that
is small, i.e. localized to essentially on one site, large,
i.e. much larger than few sites, as well as having distinct intermediate
features depending on the parameters. The Davydov soliton is a
special kind of large polaron that is formed when phonons
respond in an organized, coherent way to the presence of the
exciton.  Both the large polaron and the Davydov soliton are
approximate, semiclassical solutions since their creation involves
a large number of phonons; they are described mathematically
through the celebrated Nonlinear Schr\"odinger Equation (NLS), an
integrable nonlinear partial differential equation~\cite{dav82,dau06}.
Discrete breathers, on the other hand, are  localized solutions of
discrete nonlinear equations and differ substantially from the
extended solitons or large polarons.  They involve a local lattice
oscillation that is stable due to the disparity of its frequency
to the linearized frequency modes of the lattice.  Although
polarons, solitons and DB's are in many ways related, especially
in some limits, it is useful to differentiate among them in order
to obtain a clearer understanding of the problem.
An important mathematical
difference between polarons on one hand and solitons or DB's on
the other is that the former appear in coupled systems involving
two fields, e.g. excitons coupled to phonons, whereas the latter
are typically single field nonlinear equation solutions.
In the latter the second field has been eliminated through an
approximate procedure and replaced by an effective
nonlinear term acting explicitly on the degree of
freedom of interest.

In this review will attempt to describe the various theoretical
approaches  used in the analysis of the acetanilide study and related them
to the  different experiments.   We will start (section 2) with a brief
description
of the original infrared absorption experiment of Careri et al.
that linked for the first time the anomalous amide-I band to a
selftrapped state. In order to proceed with the theoretical analysis
we will introduce the Holstein Hamiltonian and focus first on
a semiclassical treatment (section 3).
This will enable us to describe the standard
adiabatic-like polaron acetanilide picture as
provided by Scott~\cite{sco92}.  Subsequently (section 4) we focus on a
fully quantum
mechanical approximate analytical approach that will give further insight
on the acetanilide polaron.  In section 5
we discuss the issue of the free exciton
as appears in the semiclassical as well as the approximate
quantum mechanical treatments.
Subsequently (section 6) we discuss
exact numerical solutions of the quantum
problem and point out similarities as well as differences to the
semiclassical solutions.  In section 7 we present the pump-probe
experimental results and compare them to the exact numerics.  Finally
in section 8 we summarize this work and conclude.

\section{The acetanilide story}
\label{s2}
\subsection{Acetanilide and the early experiments }
Crystalline acetanilide (CH$_3$-COONH-C$_6$H$_5$)$_n$ is an
organic molecular crystal  that forms two close hydrogen-bonded
chains running along the $b$ direction of the lattice (Fig. 1).
The nearly planar amide groups have bond distances comparable to
those in polypeptides and, as a result, it may be studied instead
of more complex protein and still provide useful polypeptide
dynamical information. The infrared absorption spectrum shown in
Fig. 2 shows two main peaks in the amide-I spectral region; a
strongly temperature dependent line at approximately
$1650~\rm{cm}^{-1}$ as well as basically temperature independent
one at $1665~\rm{cm}^{-1}$~\cite{car83,car84}. While the "normal"
$1665~\rm{cm}^{-1}$ peak was assigned to one quantum of a  C=O
vibration of an acetanilide hydrogen bonded chain, the origin of
the anomalous band was not so clear and could be a signature of a
Fermi resonance or some structural transition.  Both these
possibilities were ruled out by Careri et al. favoring a then
"unconventional" explanation that related the band to a
"Davydov-like soliton" arising through coupling of the amide
exciton to phonon modes~\cite{car83}. Careri, Scott and
collaborators used experimental indications and assumed that the
C=O excitation is coupled to an optical stretching mode of the
hydrogen bond connecting carbon and nitrogen atoms within the
peptide group~\cite{car84,eil84}. Absence of involvement of
acoustic modes in the $C=O$ selftrapping was indicated later
through  neutron scattering experiments by Barthes et
al.~\cite{bar88}. The theoretical picture that the anomalous band
is due to a dynamically localized selftrapped state permit the
quantitative explanation of the line temperature dependence using
the physics of color centers~\cite{fit68}. The  experimental and
theoretical picture that emerged in the late eighties  asigning
the anomalous spectral line  to a small Holstein-like polaron has
been corroborated by the recent experiments of Hamm et al.
~\cite{edl02a,edl02b,edl03,edl04} that will be reviewed in more
detail later. A brief historical exposition of the acetanilide
developments is given in the Appendix I.

\begin{figure}[t]
\centerline{\epsfig{figure=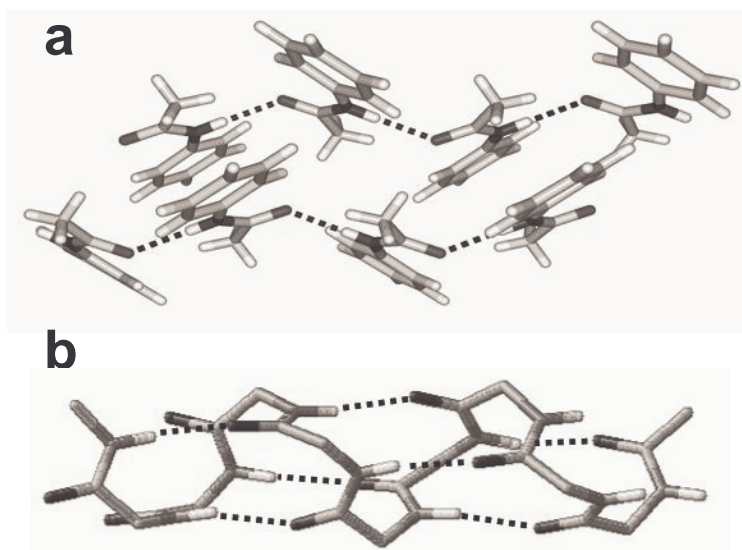,
      width=.7\columnwidth, bbllx=80 pt,
  bblly=340 pt,bburx=530 pt,bbury=730 pt, clip=}}
\caption{Comparison of (a) crystalline acetanilide; dots denote the two hydrogen bond chains  formed, and  (b) an
$\alpha$-helix that is part of a polypeptide. The $\alpha$-helix forms
three hydrogen bond chains.
} \label{fig1}
\end{figure}

\begin{figure}[tbh]
\centerline{\epsfig{figure=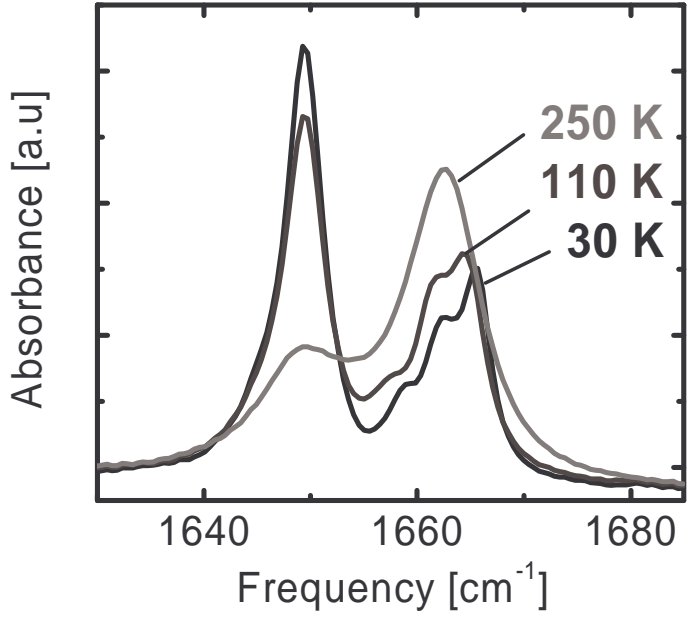,
      width=.7\columnwidth, bbllx=70 pt,
  bblly=400 pt,bburx=470 pt,bbury=730 pt, clip=}}
\caption{IR spectroscopy of acetanilide. The "normal" and the
"anomalous" peak. } \label{fig2}
\end{figure}

\subsection{The Holstein Hamiltonian}
Davydov's molecular soliton model involves the coupling of  a
vibrational exciton to acoustic phonons.  The interaction of the
excitation with phonons, modifies the latter that in-turn affect
the former in a selfconsistent way, leading, under several
approximations, to an NLS-like
soliton~\cite{dav73,dav77,dav79,dav82}.  In the case of
acetanilide, the initial indications were that the phonons
responsible for selftrapping were optical in the frequency range $
20~\rm{cm}^{-1} \leq  \omega \leq  100~\rm{cm}^{-1}$; as a result,
Careri et al. used  the Holstein Hamiltonian instead of the
Davydov one. While the theoretical work of Alexander and
Krumhansl~\cite{ale86} used coupling to acoustic phonons,
subsequent experimental work by Barthes did not indicate a strong
presence of an acoustic branch in the low ACN IR
spectrum~\cite{bar88,bar89,bar91}. Additionally, recent pump-probe
experiments seem to show that a unique phonon of frequency
approximately equal to $\omega = 50~\rm{cm}^{-1}$ is responsible
for the selftrapping phenomenon~\cite{edl02a}. Although one may
not consider the absence of an acoustic branch or even an optical
branch with dispersion as a closed issue, we may, following the
overwhelming number of approaches, make the assumption that the
physics of the amide-I lines may be described by the following
Holstein Hamiltonian~\cite{hol59a}:

\begin{eqnarray}
H=H_{ex} + H_{ph} + H_{int}\label{e1}\\
H_{ex}=\hbar \Omega \sum_{j=1}^N \left(B_J^\dag B_j + 1/2 \right)
-
 J \sum_{j=1}^N \left({ B_j}^\dag B_{j+1} + {B_j}^\dag B_{j-1} \right)
\label{e2}\\
H_{ph} =
\hbar \omega \sum_{j=1}^N \left({ b_j}^\dag b_j +1/2\right)\label{e3}  \\
H_{int}= \chi \sum_{j=1}^N {B_j}^\dag B_j
 \left({ b_j}^\dag + b_j\right) \label{e4}
\end{eqnarray}
with
\begin{eqnarray}
b_j=\frac{1}{\sqrt{2}} \left( \sqrt{\frac{m \omega }{\hbar}} q_j + \frac{i}{\sqrt{\hbar m \omega}} p_j \right) \label{e5}\\
{b_j}^\dag=\frac{1}{\sqrt{2}} \left( \sqrt{\frac{m \omega }{\hbar}} q_j - \frac{i}{\sqrt{\hbar m \omega}}p_j \right)
\label{e6}
\end{eqnarray}
where $B_j^\dag$ ( $B_j$)  creates (annihilates) a C=O vibron at
site $j$ (of $N$ sites in total),
$q_j$, $p_j$ are the position and momentum of the
$j$-th optical oscillator respectively while ${b^\dag}_j$, $b_j$
are the corresponding phonon creation and annihilation  operators.
In Eqs. (\ref{e1}-\ref{e4}) $\hbar \Omega$ is the energy of the bare
amide-I exciton equal to $\hbar \Omega \approx
1565~\rm{cm}^{-1}$ while, the optical phonon  energy is, as noted
previously, $\hbar \omega \approx 50~\rm{cm}^{-1}$. The exciton
nearest neighbor overlap is of the order $J \approx
5~\rm{cm}^{-1}$ while exciton-phonon coupling is $\chi \approx
25~\rm{cm}^{-1}$. We note that acetanilide  presents a unique problem in
molecular crystals were the three parameters entering in the
Hamiltonian, viz.  $J$, $\chi$ and $\omega$  are known to
reasonable confidence.

The C=O dipole-dipole interaction transfer integral is very small
in acetanilide, and thus, the bear exciton band is very narrow and
the exciton is immobile.  Since the phonon frequency $\omega$ is
much larger than $J$, phonons react very fast to the slow excitons
that tunnel from site to site; the phonons then follow
adiabatically the slowly tunneling exciton motion. Finally, since
the exciton-phonon coupling $\chi$ is relatively large compared to
$J$, one expects that a small, very localized polaron may form
that is not very mobile.  In order to  treat the ACN parameter
regime in the Holstein model we must resort in principle to a
fully quantum mechanical treatment since we are in the regime
where the phonon frequency is much larger than the exciton
transfer $J$. Since the first treatment was done through the
Davydov ansatz, we will  first resort to semiclassics and
subsequently investigate the connection of this approach with the
fully quantum mechanical one.

\section{Semiclassical Holstein polaron in  one dimension}
\label{s3}
\subsection{Classical phonon approach}
\label{sub3.1}
When either $J$ or $\chi$ are large many phonons are excited and
the zero point quantum motion is not important; in these cases we may use
the semiclassical limit and treat phonons
classically~\cite{kal98a}.   In many practical cases, the
excitation of even few phonons is sufficient for rendering the
modes classical.  Under this assumption, while the excitons retain
their full quantum nature, the phonon operators $b_j^\dag$ and
$b_j$ become  c-numbers denoted by $b_j^*$ and $b_j$, respectively.
Furthermore, if in this multiple phonon regime, the atomic motion
becomes more sluggish, then we may consider the now classical
phonons as "slow", leading to  $p_j \approx 0$, i.e $b_j^*
\approx b_j$, i.e. $b_j$ is real. With these assumptions, the
semiclassical Hamiltonian reads
\begin{equation}
H_{ex}=\sum_{j=1}^N \left[ \hbar \Omega   B_j^\dag B_j   -
 J ( B_j^\dag  B_{j+1} + B_j^\dag  B_{j-1} ) +
 \hbar \omega b_j^* b_j +\chi  B_j^\dag B_j
 (b_j^* + b_j ) \right] \label{e23}.
\end{equation}
In order to apply a variational procedure we need first to express
the Hamiltonian operator in a specific representation, since
otherwise operators would have to be equated with c-numbers. Using the
one-exciton state
\begin{equation}
| \psi \rangle = \sum_{j=1}^N \psi_j B_j^\dag |0\rangle_{ex}
\label{e24.1}
\end{equation}
we obtain the expected value of the Hamiltonian of Eq. (\ref{e23})
with respect to the state of Eq. (\ref{e24.1}):
\begin{equation}
H_{sc}=\langle\psi | H_{ex} | \psi \rangle =\sum_{j=1}^N \left[
\hbar \Omega  | \psi_j |^2  -
 J ( \psi_j^* \psi_{j+1} + \psi_j^*  \psi_{j-1} ) +
 \hbar \omega b_j^* b_j +\chi   | \psi_j |^2
 (b_j^* + b_j ) \right] \label{e24.2}
\end{equation}
Variational minimization of Eq. (\ref{e24.2}) with respect to the
$b_j^*$ , gives
\begin{equation}
b_j = -\chi | \psi_j |^2 \label{e24.3}
\end{equation}
leading, after substitution to the Hamiltonian (\ref{e24.2}) to
\begin{equation}
H_{sc} =\sum_{j=1}^N \left[ \hbar \Omega  | \psi_j |^2  -
 J ( \psi_j^* \psi_{j+1} + \psi_j^*  \psi_{j-1} )
-\frac{\chi^2}{\hbar \omega}   | \psi_j |^4
 \right] \label{e24.4}
\end{equation}
Use of Hamilton's equations of motion for the variables $\psi_j$ and
$i \psi_j^*$ results in the dynamical equation
\begin{equation}
i \hbar \dot{\psi_j} = \hbar \Omega  \psi_j -J(\psi_{j+1} + \psi_{j-1}) - \frac{2 \chi^2}{\hbar \omega } | \psi_j |^2 \psi
\label{e24.5}
\end{equation}
This is the celebrated Discrete Selftrapping Equation (DST) or
Discrete Nonlinear Schr\"odinger Equation
(DNLS)~\cite{eil85,kal98a}. It is a semiclassical equation for the
vibrational exciton of an amide-I mode obtained after the complete
elimination of the phonon degrees of freedom; it provides an approximate
description for the dynamics of the exciton when coupled to
phonons.

Within the classical phonon assumption, it is possible to improve
the DNLS approximation of Eq. (\ref{e24.5}) by retaining the
dynamics of the classical phonon modes.   To this effect, instead of
performing the variational minimization to Eq. (\ref{e24.1}), we may
use Hamilton's equations with respect to phonons and
obtain a coupled set of two equations, one for the exciton
amplitudes and the second for the classical phonons~\cite{kal98a}.  The
resulting set is compatible with the Born-Oppenheimer
approximation:
\begin{eqnarray}
i  \hbar \gamma \frac{d \psi_j}{d \tau} = - ( \psi_{j+1} +\psi_{j-1})+
\alpha  \psi_j u_j  \label{e24.6}\\
\frac{d^2 u_j }{d \tau^2 } + u_j = - \alpha |\psi_j |^2 \label{e24.7}
\end{eqnarray}
where $u_j$ a  dimensionless displacement  with  $\tau =
\omega t$, $\alpha = \chi \sqrt{2/\hbar \omega J}$ and  $\gamma
=\hbar \omega /J$~\cite{kal98a}.  The parameter $\gamma$ controls
the relative time scales of phonons compared to $J$.  For
stationary phonon displacements we recover the DNLS equation; this
may also  accomplished in the limit $\gamma \rightarrow 0$.

\subsection{Coherent state treatment}
\label{sub3.2}
It is instructive to use an alternative approach based on coherent
states in order to investigate the semiclassical limit of the
Holstein Hamiltonian.   We now make no assumption on the nature of
the phonons but assume that the latter are distributed in coherent
states,  viz. the individual quantum vibrational oscillators are
displaced.  As a result, we use instead of the state of Eq.
(\ref{e24.1}) the following:
\begin{eqnarray}
| \psi \rangle = \sum_{j=1}^N \psi_j {B_j}^\dag |0\rangle_{ex} |\phi_j \rangle  \label{e24.8}\\
|\phi_j \rangle = e^{\phi_j b^\dag_j -\phi^*_j b_j } |0\rangle
\label{e24.9}
\end{eqnarray}
where $\phi_j$, ${\phi^*}_j$ are the complex coefficients that
characterize the coherent state at the $j$-th oscillator and
$|0\rangle$ is the phonon vacuum.  Application of state of Eq.
(\ref{e24.8}) to the Hamiltonian (\ref{e1}-\ref{e4}), and ignoring
the constant terms arising from the zero point motions leads:
\begin{equation}
H_{sc}^\prime =\sum_{j=1}^N \left[ \hbar \Omega  | \psi_j |^2  -
 J ({ \psi_j}^* \psi_{j+1} + {\psi_j}^*  \psi_{j-1} ) +
 \hbar \omega |\phi_j |^2 +\chi   | \psi_j |^2
 (\phi_j + {\phi^*}_j ) \right] \label{e24.30}
\end{equation}
 Variational minimization of Eq. (\ref{e24.30}) with respect
to $\phi_j^*$ gives the condition
\begin{equation}
\phi_j = -\frac{\chi}{\hbar \omega} |\psi_j |^2 \label{e24.11}
\end{equation}
that, upon substitution to Eq. (\ref{e24.30}) results in the
semiclassical Hamiltonian of Eq. (\ref{e24.4}).  In other  words
this method leads to the same DNLS Eq. (\ref{e24.5}), as the
classical phonon approach.  Furthermore,  treating $\phi$, ${i
\phi^*}$, as conjugate dynamical variables and applying Hamilton's
equations to the Hamiltonian of Eq. (\ref{e24.30}) results
straightforwardly in the equations of motion (\ref{e24.6},\ref{e24.7}).

We observe that both methods used, viz. the one that treats the
phonons directly as classical and the other that is fully quantum mechanical
but assumes the phonons are in local coherent states, lead to {\it
identical} dynamical equations for the excitons.  These are the
DNLS equation, when the phonons  are completely eliminated, or the
coupled set of Eqs. (\ref{e24.6},\ref{e24.7})  in the Born-Oppenheimer
approximation.  As a result, the variational coherent state method
is simply an equivalent but different way of performing the
semiclassical approximation.
\subsection{Acetanilide analysis}
\label{sub3.3}
The semiclassical equations (\ref{e24.5}) as well
as (\ref{e24.6},\ref{e24.7})
may be used for the
analysis of the acetanilide problem.
We follow Careri et al.~\cite{car84} and
observe that DNLS equation of (\ref{e24.5}) has
two types of solutions:  (i) In the limit $N \rightarrow \infty$
the extended, free exciton
solution  $\psi_j = 1/\sqrt N e^{-i(kan-E t)/\hbar}$
with energy $E$,
wavevector $k$ and where $a$ is the lattice spacing.
The energy of this
solution is
\begin{equation}
E_{ex}(k)=\hbar \Omega -2J \cos(ka) \label{}
\end{equation}
where $k$ is the wavevector and $a$ the lattice spacing. The
extended solution describes a band of Bloch wave excitons with
very small dispersion and  bandwidth $4J$; the lowest energy state
is that of the $| k=0 \rangle $ exciton state,  the one that is
typically excited in spectroscopy with energy $E_{ex}=E(k=0) =
\hbar \Omega - 2J$. (ii) The  selftrapped solution where the
probability amplitude $\psi_j$ is centered at a given lattice site
and decays exponentially around that site. The selftrapped state
may be constructed numerically through an iterative procedure
starting from the anticontinuous limit at zero coupling, i.e for
$J=0$; it corresponds to a simple discrete breather of the DNLS
equation\cite{kal98a}. To lowest  order the selftrapped state
energy is calculated by substituting $\psi_j = \exp(-iE_{pb}
t/\hbar )$ and ignoring the inter-site coupling $J$ that is small;
we obtain $E_{pb} =\hbar \Omega -{\chi}^2/\hbar \omega $.  The
binding energy then of the selftrapped solution, i.e the total
excitonic energy gain of the selftrapped solution with respect to
the free exciton solution is
\begin{equation}
E_b =E_{ex} - E_{pb} \approx -\frac{\chi^2 }{ \hbar \omega }+2J
\label{e22b}
\end{equation}
This is approximately the amount of energy one gains by forming a
localized state in one dimension compared to the extended, long
wavelength Bloch state.

According the interpretation of  Careri
et al.\cite{car84,sco92},
the temperature dependent peak of acetanilide corresponds
to the selftrapped state, the peak at $1665~\rm{cm}^{-1}$ is the
free exciton, while the difference in energy between the two peaks
gives the selftrapped state binding energy.  We note that the
selftrapped state is a small semiclassical Holstein
polaron described trough the DNLS equation; as a result we may
also call it a discrete breather\cite{kal98a}.
As expected, this is
a fully localized state that can only acquire band character if
the eliminated quantum fluctuations are put back in the picture.
\begin{figure}[t]
\centerline{\epsfig{figure=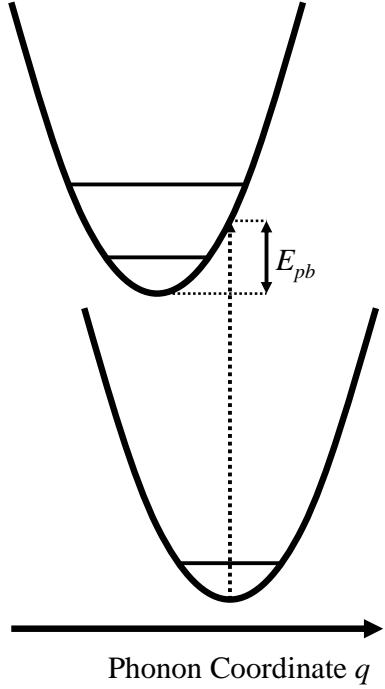,
      width=.4\columnwidth, bbllx=90 pt,
  bblly=230 pt,bburx=430 pt,bbury=780 pt, clip=}}
\caption{The displaced oscillator picture; $E_{pb}$ is the exciton energy gain
due to coupling to phonons.} \label{fig3}
\end{figure}
The semiclassical picture for the IR absorption experiments of the
amide-I mode assumes that only one
optical phonon mode of energy $\hbar \omega \approx
50~\rm{cm}^{-1}$ participates the process; this leads in the
adoption of the Holstein model for the description of the physics.
Three assumptions are subsequently made, viz. (i) the phenomena
may be described by a one dimensional model due to the quasi-one
dimensional nature of the crystal in the direction of the hydrogen
bonds. (ii) multiple phonons are excited, and as a result, the
phonon variables may be taken to be classical and (iii) these
classical phonon variables may be very slow.  The semiclassical
picture then gives a clear physical picture for the two peaks
assigning the "anomalous" one to a small polaron selftrapped
state and the normal one to a free exciton state.

Although the semiclassical picture is successful in the assignment
of the two spectral lines it cannot address the temperature
dependence of the anomalous peak; for the latter one needs to
consider directly phonon excitations, that nevertheless have been
eliminated due to the classical phonon approximation.  One way to
proceed is by ignoring polaron hoping motion altogether (since $J$
is small) and requantizing the previously assumed classical
phonons. We obtain the picture of Fig.~\ref{fig3} where the two
parabolas correspond to phonons around the no-exciton and
one-exciton states respectively; the latter state is displaced
with respect to the no-exciton state due to the exciton-phonon
interaction. In this picture, the polaron line temperature
dependence is obtained similarly as in the cases of color centers
in molecular solids\cite{fit68}. It is provided through the sum of
all overlaps between equal phonon quanta states of the ground and
displaced excited states weighted by the initial (thermal)
population of the phonon states, viz.~\cite{sco92}
\begin{eqnarray}
W(T)=\sum_{n=0}^{\infty}P_n | \langle n| \phi_n \rangle |^2 \\
P_n = \left[ 1-\exp(-\frac{\hbar \omega}{k_BT} \right] \exp(-n
\frac{\hbar \omega }{k_BT} )
\end{eqnarray}
where $ |n\rangle$ is ground state with $n$ phonons, $|\phi_n
\rangle$ is a displaced oscillator state with equal number of
phonons, while $P_n$ gives the equilibrium distribution of phonons
at temperature $T$.  The
Frank-Condon factor  $W(T)$ is found to be~\cite{sco92}:
\begin{equation}
W(T)=\exp\left[ -\frac{\chi^2}{\hbar \omega^2 }
\coth\left(\frac{\hbar \omega}{2 k_b T}\right) \right]
 I_0 \left[   \frac{\chi^2}{\hbar \omega^2 } \rm{csch}\left(\frac{\hbar \omega}{2 k_b T}\right)           \right]
\end{equation}
where $I_0$ is the modified Bessel function of first kind and
order zero. It is found that this expression
describes correctly the temperature dependence of the
anomalous peak~\cite{sco92}.

Although the semiclassical picture with the quantum mechanical
adjustment for the displaced oscillator Franck-Condon lines
appears to describe fully the ACN data, alternative explanations
and other issues have been raised in the past.  These include (i)
A Fermi resonance, topological defects or non-degeneracy in the
hydrogen-bond  may lead equally to a similar double peak feature
in the absorption spectrum. (ii) The overall validity of the
approximations involving Davydov-like adiabatic derivations may
not be justifiable. (iii) Can acetanilide be described faithfully
through a one dimensional Hamiltonians? In what regards
alternative explanations for the double peak, there is consensus
that the experimental and theoretical work performed  appears to
rule most of them out.  The experiments of Barthes et al. have
ruled out convincingly the topological defect suggestion while
also shown that the hydrogen bond non-degeneracy suggested by
Austin and coworkers is not valid~\cite{fan90,john95}. Furthermore, the analysis
of Careri and Scott as well as the recent analysis of Hamm have
shown that the occurrence of a Fermi resonance  does not seem very
likely~\cite{car84,edl02b}.  The validity of the adiabatic
approach used by Scott and collaborators has been also discussed,
however works focused more on the general Davydov soliton problem
rather that the specific acetanilide
application~\cite{brown86c,brown89}. Acetanilide is a narrow band
solid and, as a result, the occurrence or a Davydov-like extended
object is unlikely. As a result the acetanilide small polaron
regime may be better analyzed either perturbatively using $J$ as a
small parameter or through the direct numerical diagonalization.
Finally the issue of the true dimensionality of the ACN
selftrapping phenomenon demands also attention since small
polarons in one and three dimensions have quite different
properties\cite{kal98a,aub07}. These issues will be addressed in
the following sections.

\section{Quantum analysis in the small $J$ limit}
\label{s4} While the semiclassical analysis is useful as well
as intuitive, the mere fact that the hoping matrix element
$J$ is quite small in
acetanilide allows for a complete quantum mechanical solution in the
limit
$J=0$ and a subsequent use of $J$ as a perturbation
parameter~\cite{ale86,sco89}. This approach is fully
quantum mechanical and thus independent of any type of
approximations.

\subsection{Quantum $J=0$ limit}
\label{sec:1.1} When the nearest neighbor coupling is zero the
problem can be diagonalized exactly.  The Hamiltonian becomes for $J=0$:
\begin{equation}
H=
\hbar \Omega \sum_{j=1}^N \left( B_j^\dag B_j + 1/2 \right) +
\hbar \omega \sum_{j=1}^N \left( b_j^\dag b_j +1/2\right)
+\chi \sum_{j=1}^N B_j^\dag B_j
 \left( b_j^\dag + b_j\right) \label{e7}
\end{equation}
We perform the Lang-Firsov transformation~\cite{lan63} by introducing the unitary operator  $U=e^{-S}$ where
\begin{equation}
S=\frac{\chi}{\hbar \omega} \sum_{j=1}^N B_j^\dag B_j
 \left( b_j^\dag - b_j\right)  \label{e8}
\end{equation}
For the diagonalization it is useful to consider the following identities:
\begin{eqnarray}
U^\dag  B_l^\dag B_l U = B_l^\dag B_l\label{e9.1} \\
U^\dag  b_l^\dag U = b_l^\dag -\frac{\chi}{\hbar \omega} B_l^\dag B_l \label{e9.2}\\
U^\dag { b_l} U = {b_l} -\frac{\chi}{\hbar \omega} B_l^\dag B_l \label{e9.3}\\
U^\dag { b_l}^\dag b_l U =  b_l^\dag b_l -\frac{\chi}{\hbar
\omega} B_l^\dag B_l (b_l^\dag +b_l ) + \left(\frac{\chi}{\hbar
\omega}\right)^2 B_l^\dag B_l B_l^\dag B_l \label{e9.4}
\end{eqnarray}
The transformed Hamiltonian $\tilde{H}=U^\dag H U$ after removing the zero-point motion
terms  becomes
\begin{equation}
\tilde{H}=\sum_{j=1}^N\left[ \left(\hbar \Omega
-\frac{\chi^2}{\hbar \omega} \right) { B_j}^\dag B_j  -
\frac{\chi^2}{\hbar \omega}{B_j}^\dag{B_j}^\dag B_j B_j + \hbar
\omega { b_j}^\dag b_j
 \right]
\label{e13}
\end{equation}
Equation (\ref{e13}) may be diagonalized directly,
leading in the case of one vibron to the
energy spectrum:
\begin{equation}
E= \left(\hbar \Omega  -\frac{\chi^2}{\hbar \omega} \right) + n
\hbar \omega \label{e13a}
\end{equation}
for $n=0,1,2,...$. We note that the one vibron excitation has
lowered its energy due to the coupling to the phonons by an amount
equal to (see Fig.~\ref{fig3})
\begin{equation}
E_{pb}=-\frac{\chi^2}{\hbar \omega}.
\label{e14}
\end{equation}
The quantity $E_{pb}$ is the polaron binding energy determined
quantum mechanically in the $J=0$ limit.  This result coincides
with the semiclassical polaron binding energy calculated in section
(\ref{sub3.3}).

\subsection{Quantum $J \rightarrow 0$ limit}
\label{sec:1.2} When $J \neq 0$ the unitary transformation applied
to the vibron transfer term gives
\begin{eqnarray}
U^\dag H_{J} U =- J \sum_{j=1}^N \left(U^\dag  B_j^\dag B_{j+1} U +
U^\dag B_j^\dag B_{j-1} U \right) =\nonumber\\
- \sum_{j=1}^N \left( J_{j,j+1}  B_j^\dag B_{j+1}  +
J_{j,j-1} B_j^\dag B_{j-1} \right)\label{e16}\\
J_{j,j+1} = J\exp \left\lbrace  -\frac{\chi}{\hbar \omega}
 \left[  ( b_j^\dag  - b_j ) - (b_{j+1}^\dag - b_{j+1} )
\right]  \right\rbrace
\label{e17}
\end{eqnarray}
In order to proceed we assume first that the system is at zero
temperature and the phonons at  each site are in their
corresponding ground state, viz.  $|0 \rangle \equiv| 0 \rangle_{ph} =
... |0\rangle_{j-1} |0\rangle_j |0\rangle_{j+1} ...$;  using this
state we find an effective coupling between
adjacent sites:
\begin{equation}
J_{eff}=\langle 0 |J_{j,j+1} | 0\rangle = J \exp\left[
-\left(\frac{\chi}{\hbar \omega }\right)^2 \right] \label{e18}
\end{equation}
The effective bandwidth of the polaron tunneling is
thus reduced as a result of the interaction with phonons~\cite{hol59b}.
This renormalization of the polaron bandwidth is an approximate, mean
field result, that ignores local phonon fluctuations and has a
small effect when  $J$ is small ($J \ll \hbar \omega$). While
the Lang-Firsov transformation does not depend on the
value of $J$, the
averaging over the zero phonon states should be done for small hopping
matrix elements.

Using  Eqs. (\ref{e18}) we may now write the transformed
Hamiltonian in the $J\rightarrow 0$ limit for the case of zero
phonons as follows:
\begin{equation}
\tilde{H}=\sum_{j=1}^N\left[ (\hbar \Omega  -\frac{\chi^2}{\hbar
\omega} ) { B_j}^\dag B_j  - J_{eff} \left( {B_j}^\dag B_{j+1}
+{B_j}^\dag B_{j-1} \right) - \frac{\chi^2}{\hbar
\omega}{B_j}^\dag{B_j}^\dag B_j B_j
 \right] \label{e19}
\end{equation}
 If we assume that only one $C=O$ vibron is present while
phonons are in their ground state, i.e.
\begin{equation}
| \psi \rangle = \sum_{j=1}^N \psi_j {B_j}^\dag |0\rangle_{ex} |0
\rangle_{ph} \label{e20}
\end{equation}
then we can directly diagonalized the tight-binding Hamiltonian. Using
Eq. (\ref{e20}) in Eq. (\ref{e19}) leads to
\begin{equation}
E \psi_j = \left(\hbar \Omega  -\frac{\chi^2}{\hbar \omega}
\right) \psi_j -J_{eff} \left( \psi_{J+1} + \psi_{j-1} \right).
\label{e21}
\end{equation}
The energy spectrum for the one-vibron sector is
\begin{equation}
E(k) = \left( \hbar\Omega -\frac{\chi^2}{\hbar \omega } \right)
-2Je^{-(\frac{\chi}{\hbar \omega})^2} \cos(ka) \label{e22a}
\end{equation}
The approximate small-$J$ procedure leads to a small polaron band
that has gained energy $-\chi^2 / \hbar \omega$ with respect to the
center of the bear exciton band (obtained for $\chi = 0$)  and
has a reduced bandwidth equal to
$4Je^{-(\frac{\chi}{\hbar \omega})^2}$.  This polaron band is the
only one-exciton, zero-phonon solution in the $J \rightarrow 0$
limit.  The approximate energy $E(k)$ for the small polaron band
of Eq. (\ref{e22a}) is identical to the one obtained by Scott using
degenerate perturbation theory~\cite{sco92}.

Let us briefly summarize the findings of the approximate but
fully quantum mechanical treatment: In the search for the
eigenstates of the Holstein Hamiltonian, the exact quantum
mechanical calculation at $J=0$ results in an eigenstate that has energy
lower by an amount equal to $E_{bp} =-\chi^2/\hbar \omega$ from
the bear exciton state.  This state is thus favored energetically over the
bear exciton that  exists when $\chi = 0$. The value
of the binding energy of this new
polaron state coincides with the semiclassical (DNLS) binding energy.
For exciton hoping $J$ small we find
(at $T=0$) that
this state acquires dispersion with bandwidth $B=4J \exp
\left[ -(\chi/\hbar \omega )^2 \right] $.  Consequently, this
solution provides the small polaron {\it band} formed as result of
exciton-phonon coupling and the small $J$ value.  Since the
semiclassical and quantum mechanical polaron binding energies are identical,
we may consider that the localized semiclassical polarons begin to tunnel
when quantum fluctuations are included to  low order and form a band
with bandwidth $B$.  We point out that for acetanilide
$B\approx16\exp(-1/4)~\rm{cm}^{-1}~\approx
12.4~\rm{cm}^{-1}$, i.e. the bandwidth reduction due to
exciton-phonon coupling is small.

Although the polaron solutions in the semiclassical and quantum
treatments coincide, the free exciton solution found in the
context of DNLS seems to be absent in the quantum mechanical
analysis. It is in fact easy to see that in the fully quantum
approach the exact free exciton state is not an eigensolution in
any dimension since any extended exciton wave state generates
phonons through the exciton-phonon coupling term in the
Hamiltonian. Since the presence of the free exciton peak is
central to the explanation of the acetanilide amide I spectrum one
must find a way to bypass this problem.  One possibility that is
compatible with the semiclassical picture is to consider that the
free exciton state is actually an excited polaron state that
involves a number of excited phonons. However since the phonon
involved in the process has energy equal to $50 cm^{-1}$ while the
polaron binding energy is less than half of this value
(approximately $16 cm^{-1}$), an excited polaron plus one phonon
cannot match the value of $16 cm^{-1}$, except for exceptionally
large values of $\chi$ that are not reasonable for molecular
crystals. Another possibility is that the description of
acetanilide as a one dimensional solid is not sufficient.  It is
known semiclassically that in $3D$ the polaron state may be
separated by a barrier from the free exciton state and, as a
result, both states may coexist in some fashion\cite{kal98a}.  If
such a picture survives in the fully quantum case and for the
proper parameter regime, it  might then be possible to identify
the two states accordingly.  These issues will be addressed in
detail later.  We note here that the small-$J$ analysis can be
done easily in three dimensions leading to a small polaron band
with energy
\begin{equation}
E({\bf k}) = \left( \hbar\Omega -\frac{\chi^2}{\hbar \omega } \right)
-2e^{-(\frac{\chi}{\hbar \omega})^2} \left[ J_x \cos(k_x a_x) +
J_y \cos(k_y a_y) +J_z \cos(k_z a_z) \right]
\end{equation}
where
${\bf J} \equiv (J_x , J_y , J_z )$ are the hopping rates for the
three Cartesian directions respectively,
${\bf k}=(k_x , k_y , k_z )$ is the wavevector and ${\bf a} = (a_x , a_y , a_z )$ the lattice spacings.

\subsection{Brief summary on various approaches}
It is worthwhile to summarize briefly the various equations
obtained so far through three different approaches.
\begin{itemize}
\item The semiclassical DNLS equation may be obtained
variationally either by (a) considering phonons classical or (b)
assuming that phonons are quantum but distributed according
to coherent states. In both cases we obtain
identical results, viz. energy gain for the polaron and no bandwidth reduction.
In the Born-Oppenheimer approximation both approaches lead to
identical dynamical equations as well.

\item When we solve the Holstein model fully quantum mechanically
but in the small hopping limit we find a polaron band with energy
gain and a reduced bandwidth.
\end{itemize}
Comparing these approached we find
that all three of them (pure semiclassical, phonons in coherent states as well
as quantum mechanical) give the same polaron energy gain, yet, the bandwidth
reduction does arise only from the fully quantum mechanical approach.
Furthermore, the free exciton state is a stationary  solution
only semiclassically.
While the semiclassical approaches  result to an effective
nonlinear equation of motion, viz. DNLS, the approximate quantum
methodology leads to a
QDNLS Hamiltonian~\cite{sco94}.  The latter provides
a fully quantum mechanical
reduced description for the excitons provided the hopping rate $J$ is
small and phonons are not excited on average from their ground states.

From Eq. (\ref{e19}) we
see that the on site term of the QDNLS Hamiltonian
contains already the polaron binding
energy shift while the hopping term is also renormalized.  When
only one exciton is present, this Hamiltonian describes a small
polaron band. When more exciton quanta are present, it may
describe multiexciton states as well as interaction among them.
It is clear that the QDNLS approach is valid when intersite phonon correlations
are not important.

\section{The free exciton state in the  semiclassical analysis}
\label{s5}
In order to focus deeper into the double peak acetanilide
structure we may investigate in more detail the semiclassical
picture before ultimately probe directly the fully quantum
regime using exact numerics.  In this section we will
look into the free exciton state both in the Born-Oppenheimer
approximation in one dimension  but also treating the three
dimensional case.

\subsection{Born-Oppenheimer approximation}

\begin{figure}[t]
\centerline{\epsfig{figure=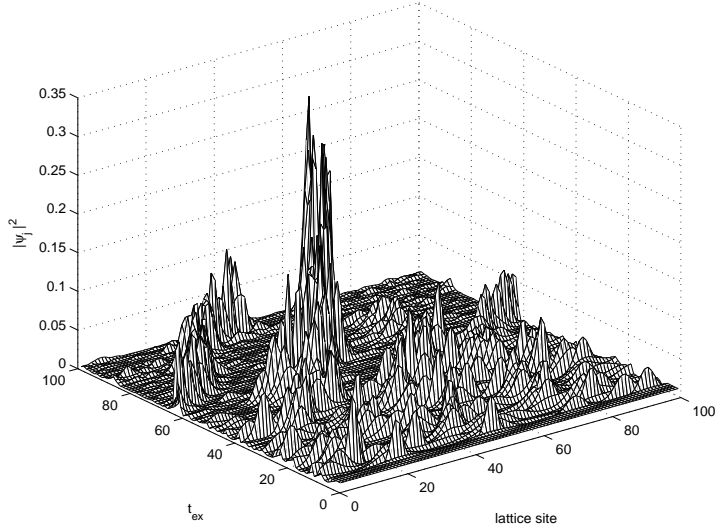,
      width=.7\columnwidth, bbllx=10 pt,
  bblly=190 pt,bburx=560 pt,bbury=600 pt, clip=}}
\caption{Time evolution  of the site probability $|\psi_j $  as a
function of  exciton time
$t_{ex} =  \tau /5 \gamma \equiv J t/5$ for $\gamma
=10$ and $\alpha =2.3$.  We used a periodic one dimensional
lattice of $100$ sites.  The initial condition is that of a free
exciton of zero momentum and uniform noise in the initial
(dimensionless) phonon
displacements  $u_j$'s with amplitude of the order of $10^{-3}$
(the phonon lattice evolution is not shown).  The free exciton is
initially stable but subsequently collapses to localized states.
\label{figg4}}
\end{figure}

In the fully adiabatic picture the classical Einstein oscillators
of the Holstein Hamiltonian have no dynamics but follow the
exciton motion.  We may still consider these oscillators classical
but allow them a characteristic time scale for response to the
exciton motion.  In this case we need to solve the  set of
equations (\ref{e24.6},\ref{e24.7}) that determine the stability of the
adiabatic solutions. The  acetanilide values correspond to
$\alpha \approx 2.3$ and $\gamma \approx 10$; thus if we
substitute the semiclassical adiabatic solutions for the small
polaron and free exciton of section (2.3) and use Eqns.
(\ref{e24.6},\ref{e24.7}) we will find what is the regime of stability of
these solutions.  The adiabatic polaron state may be constructed
via a numerically exact procedure from the anticontinuous limit;
its linear stability has already  been discussed~\cite{kal98a}.
The stability of the free exciton state, on the other hand, may be
assessed numerically by direct substitution of the state in the
equations of motion of Eqs. (\ref{e24.6},\ref{e24.7});
the result is shown in
Fig. (\ref{figg4} ). In order to engage the vibrational lattice to the free
exciton we have added initially a small component of uniform noise
in the classical phonons.  In the figure we show  the free exciton
evolution in time units $t_{ex}=\gamma \tau /5 \equiv J t /5$;
we observe that after
few exciton oscillations the free exciton becomes unstable and
forms localized structures.  The phenomenon of the semiclassical
polaron formation from the free exciton state is controlled by a
characteristic polaron formation time $\tau_p$.   This waiting
time  for  the onset of localization depends strongly on the
specific parameter regime as well as initial amplitude to the
phonon excitation.  Extensive simulations show that $\tau_p
\rightarrow \infty$ when $\gamma \rightarrow 0$, i.e. when we
address the fully adiabatic regime.
In the opposite limit of $\gamma > 1$, on the other hand,
the semiclassical free exciton becomes very
quickly unstable and localizes. We expect that as the temperature
of the system increases, this effect becomes more dominant leading
to a collapse of the free exciton into localized states.

The use of a finite parameter $\gamma = \hbar \omega
/J$  in the semiclassical problem  introduces a new time scale in
the Holstein Hamiltonian; while $\tau$ is the natural time scale
for the phonon evolution (measured in phonon periods) , the
exciton time is measured in $\tau_{ex} = \gamma \tau$.  In the
adiabatic regime $\gamma \ll 1$ and in the course of many exciton
oscillations there are very few phonon circles.  Using in Eqs.
(\ref{e24.6},\ref{e24.7})
as initial condition a free exciton state, we observe
that the slow phonon change simply modifies the local site
energies of the exciton equation; as a result the free exciton
suffers very weak phonon perturbation and survives for long times.
In the opposite regime where $\gamma \gg1$, many phonon
oscillations occur in few exciton ones; as a result the
interaction term acts as an effective multiplicative noise term
(for some initial phonon conditions) leading to a very rapid
collapse of the free exciton into localized states.  This purely
classical effect is enhanced in the fully quantum regime through
the zero point motion.

Since the parameter regime of acetanilide is in
the range $\gamma >1 $ we
find that the free exciton is generally unstable to even quite
small perturbations.   This behavior poses a problem to the
standard one dimensional adiabatic explanation since, for this
approach to be valid the free exciton state should be stable
as is the small polaron.
One way
to proceed within  the semiclassical context is to consider the
possibility that the selftrapping phenomenon takes place in three
dimensions. Since, as noted earlier, in 3D a barrier separates
free exciton and polaron, it may be that the barrier helps
in stabilizing both states.
~\cite{kal98a}. We
will proceed with the semiclassical analysis of the 3D Holstein
model in the next subsection.

\subsection{Semiclassical 3D problem}
\label{sec:2.1}

\begin{figure}[t]
\centerline{\epsfig{figure=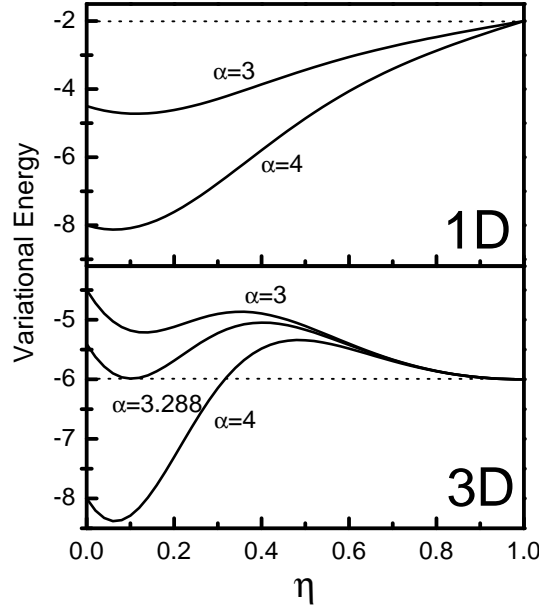,
      width=.5\columnwidth, bbllx=10 pt,
  bblly=260 pt,bburx=460 pt,bbury=770 pt, clip=}}
\caption{Variational energy in 1D (top) and 3D (bottom) for
various values of the parameter
$\alpha\equiv\sqrt{2\chi^2/\hbar\omega J}$. Adapted from
Ref.~\cite{kal98a}.\label{figp1}}
\end{figure}

It was  pointed out recently that the dipole-dipole interaction
responsible for the nearest-neighbor transfer of excitons in ACN
extends not only along the hydrogen-bonded chain direction but
also in the plane perpendicular to it\cite{ham06a};  this point will be
detailed in sections (\ref{sec:p2}) and (\ref{sec:p1}). In order to perform
the semiclassical analysis for the general problem in three
dimensions we write the semiclassical Holstein Hamiltonian as follows:

\begin{eqnarray}
H_{sc} =\sum_{\bf j} \left[ \hbar \Omega  | \psi_{\bf j} |^2  -
(\Delta \psi )_{\bf j}
 +  \hbar \omega b_{\bf j}^* b_{\bf j}
 +\chi  | \psi_{\bf  j} |^2  ( b_{\bf j}^* + b_{\bf j}  ) ~
 \right] \label{e5.1}\\
(\Delta \psi )_{\bf j}=
 J_x \left({ \psi^*_{j_x , j_y , j_z}} \psi_{{j_x }+1, j_y , j_z} +
\psi^*_{j_x , j_y , j_z}  \psi_{{j_x} -1, j_y , j_z} \right) \nonumber\\
 J_y \left( \psi^*_{j_x , j_y , j_z} \psi_{j_x , {j_y }+1,
j_z} + \psi^*_{j_x , j_y , j_z}  \psi_{j_x , {j_y} -1, j_z} \right)
  + J_z \left( \psi^*_{j_x , j_y , j_z} \psi_{j_x , j_y , j_z +1} +
\psi^*_{j_x , j_y , j_z}  \psi_{j_x , j_y , j_z -1}  \right)
\label{e5.1a}\\
\end{eqnarray}
where ${\bf j}= (j_x , j_y , j_z )$ and $J_x , J_y , J_z $ are the
transfer rates in the three perpendicular axis directions.  We
assume that the hydrogen bonded axis coincides with the $x$-axis ;
the parameters $J_x$, $J_y$, $J_z$ may take positive (negative)
values leading to negative (positive) hopping rates.
We may now perform the variational minimization with
respect to the phonon c-numbers $b_{\bf j}^*$, $ b_{\bf
j}$ and obtain the variational energy $E_v$:
\begin{eqnarray}
E_v = \sum_{\bf j} \left[ \hbar \Omega  | \psi_{\bf j} |^2  -\frac{\chi^2}{ \hbar \omega}  | \psi_{\bf  j} |^4 \right. \nonumber\\
\left. - J_x  \left( | \psi_{{j_x }+1, j_y , j_z} - \psi_{\bf j}
|^2 -2\right)  -J_y \left(  | \psi_{j_x , {j_y }+1 , j_z} -
\psi_{\bf j} |^2 -2\right) -J_z \left(  | \psi_{j_x , j_y , {j_z
}+1} - \psi_{\bf j} |^2  -2\right)\right]  \label{e5.2}
\end{eqnarray}
where $d$ is the dimensionality.  In order to probe into the
variational energy $E_v$ we make the ansatz
\begin{equation}
\psi_{\bf j} = A {\zeta}^{|j_x |} \eta^{|j_y |} \theta^ {|j_z |}
\label{e5.3}
\end{equation}
where $|\zeta |\leq  1$, $|\eta | \leq 1$, $|\theta | \leq 1$. The
three parameters $\zeta$, $\eta$, and $\theta$ take positive (negative)
values when the hopping overlap in the corresponding direction is
negative (positive).  The normalization
factor is found to be
\begin{equation}
A=\left( \frac{1-\zeta^2}{1+\zeta^2} \right)^{1/2}
\left( \frac{1-\eta^2}{1+\eta^2} \right)^{1/2}
\left( \frac{1-\theta^2}{1+\theta^2 }\right)^{1/2}
\label{e5.4}
\end{equation}
Substitution of the trial function of Eq. (\ref{e5.3})
into Eq. (\ref{e5.2}) we obtain the following expression for the variational energy
\begin{equation}
E_v = -4\left[ J_x \frac{\zeta}{1+\zeta^2}+J_y
\frac{\eta}{1+\eta^2}+J_z \frac{\theta}{1+\theta^2}\right] -
\frac{\chi^2}{ \hbar \omega}\frac{(1-\zeta^2 )(1+\zeta^4
)}{(1+\zeta^2 )^3} \frac{(1-\eta^2 ) (1+\eta^4 )}{(1+\eta^2 )^3}
\frac{(1-\theta^2 )(1+\theta^4 )}{(1+\theta^2 )^3} \label{e5.5}
\end{equation}
In the fully symmetric case of $J_x = J_y =J_z \equiv J >0$
we have $\zeta = \eta = \theta \equiv \eta$ and
obtain the general expression $d$ dimensions\cite{kal98a}
\begin{equation}
E_v = -4 d J  \frac{\eta}{1+\eta^2}-
\frac{\chi^2}{ \hbar \omega}
\frac{(1-\eta^2 )^d (1+\eta^4 )^d}{(1+\eta^2 )^3d}
\end{equation}
Figure ~\ref{figp1} shows the variational energy (in units of $J$)
in 1D and 3D for the isotropic case
with all negative hopping overlaps ($ J > 0$).
In 1D, the curve is barrierless for any set of parameters,
regardless how small the coupling is. That is, a partially
localized solution $\eta<1$  has always smaller
energy than the extended solution $\eta=1$, and furthermore, this
solution can be reached without having to surmount any barrier.
The behavior in 3D is drastically different.
For couplings below the  critical threshold value
 $\alpha_c\equiv\sqrt{2\chi_c^2/\hbar\omega J}=3.288$,
the extended solution with $\eta=1$ is the lowest energy solution.
For couplings larger than this critical value $\alpha_c$, a
strongly localized state $\eta\ll 1$ becomes the lowest energy
solution\cite{kal98a}.


\subsection{Polaron and free exciton coexistence in three dimensions}
\begin{figure}[t]
\centerline{\epsfig{figure=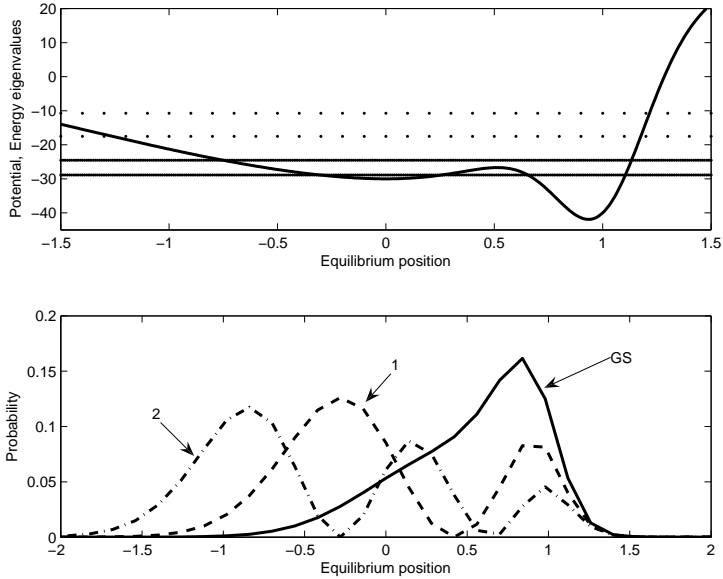,
      width=.7\columnwidth, bbllx=40 pt,
  bblly=190 pt,bburx=550 pt,bbury=610 pt, clip=}}
\caption{(a) The adiabatic potential for the 3D polaron as a
function of displacement; the horizontal lines correspond to the
first two energy eigenvalues while dots the next two
 for $\alpha =4$ and $m^* =5$. In
(b) the spatial distribution of the probability for the lowest
three eigenfunctions, the ground state (GS), first (1) and second
(2) excited states, respectively.\label{f6}}
\end{figure}

The effective semiclassical barrier between the
free exciton and small polaron forms
when $\alpha =  \alpha_c  \approx 3.288$. When $\alpha < \alpha_c$
the minimum energy state is the free exciton while for $\alpha > \alpha_c$
is the small polaron.  When the parameter values are such that
an adiabatic double well structure is formed, the transition from free
exciton to small polaron occurs through quantum lattice
fluctuations.  A simple way to include the latter while retaining
the intuitive adiabatic picture is by treating $\eta$ as  an
effective  "configurational variable" and quantizing the double
well potential problem.  In other words, we may consider that the
dynamics of the transition between polaron and exciton may be
described by a particle of effective mass $m^*$ that is moving in
the adiabatic potential.   Upon diagonalization of this
effective one dimensional problem we may obtain the states
involved and understand their dynamics. The results
are shown in Fig. ( \ref{f6} ) for  $\alpha =4$ where
we portray the adiabatic potential, the four lowest energy eigenvalues
as well as the three lowest eigenfunctions.  The equilibrium position
is zero for the undisplaced free exciton state.  For the
parameter regime shown the lowest potential minimum corresponds to
the small polaron while the smooth, wide secondary minimum is the free
exciton minimum.

From the numerical solution of the spectral problem we find that
the two lowest states are relatively close in energy and form a
type of doublet. The  ground state wavefunction has a single peak
that appears to have two components.  While it is centered close
to the polaron minimum it has a considerable overlap with the
exciton minimum as well. The first excited state, on the other
hand, is doubly-humped and centered in the wide, shallow secondary
minimum of the free exciton but with significant amplitude in the
polaron minimum.  The ground state-excited state doublet has
distinct polaron-like and exciton-like structure respectively. If
the $3D$ Holstein model is a good representation of the
acetanilide problem, then it is possible that these are the two
states that are relevant for the explanation of the double peak in
the ACN absorption spectrum.

In the previous analysis we used a relatively large value of the
nonlinearity parameter so that we are withing the double well
regime of the adiabatic potential. This choice may be justified in
the following way: Since we know that in reality the polaron
hopping rate is reduced we may in  an ad-hoc fashion use   the
effective rate of Eq. (\ref{e18}) instead of just $J$ and obtain a
new nonlinearity parameter $\alpha_{eff} = \exp \left[ \frac{1}{2}
(\frac{\chi}{\hbar \omega })^2 \right] \alpha$ that is larger than
$\alpha$. This way the effective nonlinearity increases for the
same acetanilide values  and the system is found to be closer to
the adiabatic double well regime. Clearly further investigation on
this approach is necessary.

\section{Numerically Exact Diagonalization of the Holstein Hamiltonian}
\label{sec:p2}

With recent advances in computer technology it became possible to
diagonalize the Holstein Hamiltonian on a numerically exact level,
even in the 3D case~\cite{bon99,ku02}. This allows one to study to
what extent the conclusions from the DNLS (e.g. the existence of a
barrier in the 3D case) manifest itself in the full quantum
regime. To that end, a concept introduced by Trugman and coworkers
is used~\cite{bon99,ku02}. In brief, Bloch states with momentum
$k$ are constructed:
\begin{eqnarray}
\Psi=\frac{1}{\sqrt N}\sum_{j} \Phi e^{ikj} \label{eqbasis1}
\end{eqnarray}
with site states $\Phi$ that are identical for each site $j$. The
site states are expanded in a basis $\varphi_{\{n_i\}}$:
\begin{eqnarray}
\Phi=\sum_{\{n_i\}}c_{\{n_i\}}\varphi_{\{n_i\}} \label{eqbasis3}
\end{eqnarray}
with (in one dimension):
\begin{eqnarray}
\varphi_{\{n_i\}}=|1_0;...,n_{-1},n_0,n_{+1},...\rangle
\label{eqbasis2}
\end{eqnarray}
where the '$1_0$' denotes that we have one quantum in the exciton
coordinate at site 0 (since the Hamilton is quantum-conserving
with respect to the exciton coordinate, it is block-diagonal and
we consider here only the one-quantum manifold of states). The
$n_i$ are the numbers of phonon excitations at site $i$ around the
exciton at site 0.

The essential trick of Refs.~\cite{bon99,ku02} is an efficient
strategy to select basis states, which we review here only very
briefly. We start from a 'root' basis state without any phonon
excitation: $|1_0;...,0,0,0,...\rangle$. For each generation
(total number of generations: $N_h$), basis states are added
corresponding to jumps along one of the off-diagonal elements of
the Hamiltonian Eq.~(\ref{e2}--\ref{e4}) in a tree-like manner.
This can either be one phonon jump with a matrix element
\begin{eqnarray}
\langle
1_0;...,n_{-1},n_0,n_{+1},...|H|1_0;...,n_{-1},n_0+1,n_{+1},...\rangle=-\sqrt{n_0+1}\chi
\end{eqnarray}
or a shift along one of the crystal directions with matrix
elements
\begin{eqnarray}
\langle
1_0;...,n_{-1},n_0,n_{+1},...|H|1_0;...,n_{-2},n_{-1},n_{0},...\rangle=J
\exp(ika).
\end{eqnarray}
In optical spectroscopy, where a $k=0$ selection rule exists, the
phase factor $\exp(ika)$ disappears. The basis gets a pyramidal
shape with maximum $N_h$ phonon excitations at site 0 to one
phonon excitation at one site $\pm (N_h-1)$ in either lattice
direction. The basis is infinite with respect to lattice
transformations and fully accounts for the translational symmetry
of the crystal (Eq.~\ref{eqbasis1}). However, the basis is finite
with respect to the phonon-exciton distance, and the phonons are
confined to the exciton by construct within $\pm (N_h-1)$. The
number of basis states scales as $O\left((D+1)^{N_h}\right)$,
where $D$ is the dimensionality of the system. In 3 dimensions
with generation number $N_h=11$, this reveals as many as
$1.4\times10^6$ basis states. Nevertheless, the matrix is sparse
and can be partially diagonalized using a Lanczos algorithm. On an
AMD Opteron processor with 12~GB main memory, the ground state is
calculated within minutes, or, if the whole absorption spectrum is
needed, within typically a day.

Before we start, one comment is in order: In order to distinguish
a free-exciton from a polaron solution, we will use the phonon
displacement $q$ as a criterion. It might seem more logical to
investigate the extension of a solution in real space, similar to
in the DNLS case (Sec.~\ref{sec:2.1}). However, even though the
free-exciton would, of course, be perfectly delocalized in real
space, any polaron solution would be as well in the full-quantum
case and one could not distinguish between polaron and exciton in
this way. This is by construct of the Bloch ansatz
Eq.~(\ref{eqbasis1}), which accounts for the translational
symmetry of the problem. In the strict meaning of the word, there
is no true, spatial self-trapping in the full-quantum case. This
is fundamentally different from the semi-classical DNLS case,
which does break the translational symmetry of the problem and
reveals solutions that are localized in real space. Only in the
limit of infinite effective polaron mass, where the
polaron-dispersion relation would becomes infinitely flat, one
could construct spatially localized wavepackets, which then would
indeed be eigenstates of the polaron Hamiltonian. Comparison with
Eq.~(\ref{e22a}) shows that a flat dispersion relation is obtained
for $\chi/\hbar\omega\rightarrow\infty$.

\begin{figure}[t]
\centerline{\epsfig{figure=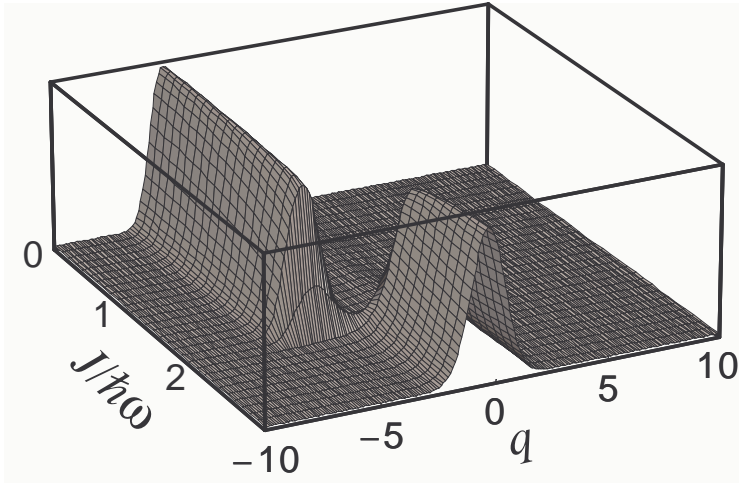,
      width=.7\columnwidth, bbllx=80 pt,
  bblly=470 pt,bburx=510 pt,bbury=760 pt, clip=}}
\caption{The Holstein Hamiltonian in 3D with isotropic negative
exciton coupling: Projection of the ground state wavefunction onto
the phonon-coordinate $q$. The exciton-phonon coupling was set to
$\chi/\hbar\omega=3.5$ in this calculation, and the exciton
coupling $J/\hbar\omega$ was varied from 0 to 3.}\label{figp2a}
\end{figure}

Fig.~\ref{figp2a} shows the results of a numerically exact matrix
diagonalization of the Holstein Hamiltonian for exciton-phonon
coupling $\chi/\hbar\omega=3.5$ with the exciton coupling
$J/\hbar\omega$ varied from 0 to 3. Shown is the projection of the
corresponding ground state wavefunction onto the phonon coordinate
$q$. With this set of parameters, the system should be close to
the DNLS regime ($\chi/\hbar\omega\gg1$), so one expects that one
can directly compare results from Sec.~\ref{sec:2.1} with that
from the exact diagonalization.

Indeed, for small exciton phonon couplings $J$, which corresponds
to a large $\alpha\equiv\sqrt{2\chi_c^2/\hbar\omega J}$, the
wavefunction is displaced with respect to the phonon coordinate,
close to the value $\sqrt 2\chi/\hbar\omega$ expected from the
displaced oscillator picture (Sec.~\ref{sec:1.1}). Referring to
Fig.~\ref{figp1}, this corresponds to a polaron solution with
small $\eta$ for $\alpha>\alpha_c$, which in this parameter range
constitutes the lowest energy solution. However, as the exciton
coupling is increased to $J/\hbar\omega\approx2.2$, which for the
given exciton-phonon coupling relates to the critical value
$\alpha_c=3.288$, the ground state wave function jumps almost
suddenly back to $q\approx0$. From Fig.~\ref{figp1} we expect for
$\alpha<\alpha_c$ that the extended state with $\eta=1$ becomes
the lowest energy solution, and accordingly, the
phonon-displacement $q$ vanishes at this point. If the system were
truly classical, the switching between the two regimes would occur
instantaneously, but since the quantum mechanical wavefunction
delocalizes, a relatively small parameter regime exists where
polaron and exciton solutions coexist (similar to Fig.~\ref{f6}).
However, we note that even though exciton-like solutions are found
in the quantum case, they are not exactly free excitons (the
displacement $q$ is not exactly zero). The free exciton is not an
eigenstate of the Holstein Hamiltonian.

\begin{figure}[t]
\centerline{\epsfig{figure=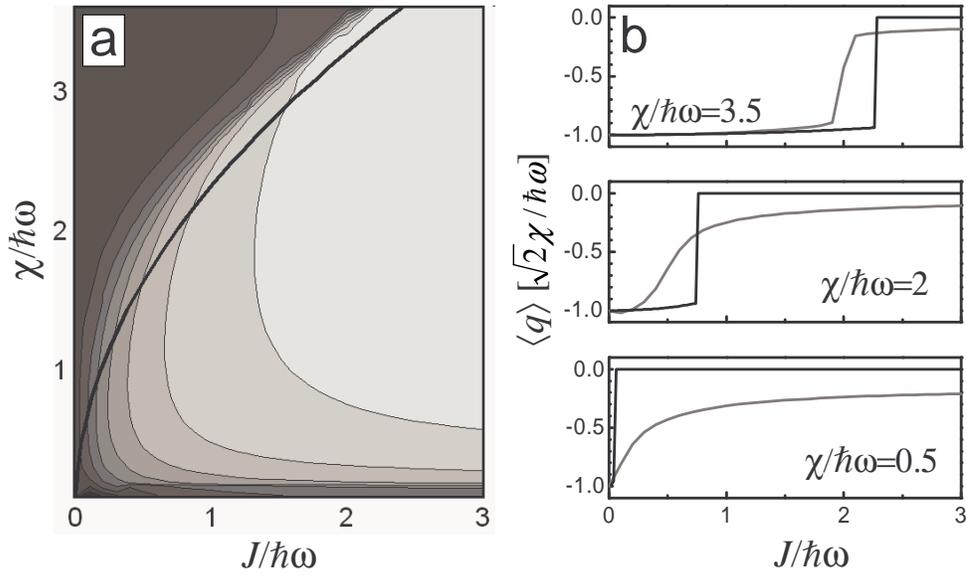,
      width=.9\columnwidth, bbllx=60 pt,
  bblly=440 pt,bburx=520 pt,bbury=720 pt, clip=}}
\caption{The Holstein Hamiltonian in 3D with isotropic negative
exciton coupling: (a) Displacement of the phonon coordinate $q$
(in units of $\chi/\hbar\omega$) of the ground state according to
a full-quantum diagonalization of the Holstein Hamiltonian.
Results are plotted as a function of exciton coupling $J$ and
exciton-phonon coupling $\chi$, both in units of the phonon
frequency $\hbar\omega$. The thick line separates the two regimes
predicted from the semi-classical DNLS:
$\alpha_c=\sqrt{2\chi_c^2/\hbar\omega J}=3.288$ (b) Horizontal
cuts through (a), exemplifying the quantum case (bottom,
$\chi/\hbar\omega=0.5$), the close to classical case (top,
$\chi/\hbar\omega=3.5$), and an turn-over regime (middle,
$\chi/\hbar\omega=2.0$). The grey lines represent the results from
an exact diagonalization, the black lines that from the DNLS,
interpreting the interaction term in Eq.~(\ref{e24.4}) as an
effective coupling $\chi_{eff}=\chi|\psi|^2$; hence the maximum
displacement found in the peak of the wavefunction is $q=\sqrt
2\chi|\psi_{max}|^2/\omega$. In either case, the displacement $q$
is plotted in units of $\sqrt2 \chi/\hbar\omega$.}\label{figp2}
\end{figure}

Fig.~\ref{figp2}a puts these results into a more general
perspective. Shown is the expectation value of the phonon
displacement $\langle q\rangle$, relative to that expected from
the $J=0$ case $q_{max}=\sqrt2\chi/\hbar\omega$, as a function of
the two parameters of the theory $\chi/\hbar\omega$ and
$J/\hbar\omega$. The thick line separates the two regimes of the
semi-classical DNLS, defined by
$\alpha_c\equiv\sqrt{2\chi_c^2/\hbar\omega J}=3.288$: Left from
that line, the DNLS predicts polaron solutions with displacements
close to the maximal value $q_{max}=\sqrt2 \chi/\hbar\omega$,
whereas extended states with $\eta=0$ are expected on the right
side of this phase diagram. For larger exciton-phonon coupling,
$\chi/\hbar\omega>\approx2$ the quantum solution essentially
follows that line, except that the turn over between polaron and
exciton solution is not discontinuous (as discussed above).
Nevertheless, as $\chi/\hbar\omega$ is increased from 2 to 3.5
(Fig.~\ref{figp2}b middle and top), the transition occurs more and
more abruptly.

In contrast, for $\chi/\hbar\omega<\approx1$ the quantum solution
deviates strongly from the DNLS prediction. In particular, polaron
solutions with large displacements $\langle q\rangle$ may exist
even for relatively large exciton coupling $J$, where the DNLS
would strictly reveal an exciton solution (Fig.~\ref{figp2}b,
bottom). This difference between semi-classical and quantum result
can be rationalized by taking into account the second term of
Eq.~(\ref{e22a}),
\begin{eqnarray}
-2J \exp\left[-\chi^2/(\hbar\omega)^2\right]\cos(ka)
\end{eqnarray}
which may be considered a quantum-correction factor relative to
the polaron binding energy in the semiclassical DNLS case
(Eq.~\ref{e13a}). The origin of the quantum-correction factor is
exciton tunneling that occurs when a spatially localized solution
(Sec.\ref{sec:1.2}) starts to delocalize due to a non-zero exciton
coupling $J\ne 0$. In that case, a phonon displacement follows the
exciton as the latter hops from one site to the next. In fact, the
factor $\exp(-\chi^2/(\hbar\omega)^2)$ equals the overlap integral
(Franck-Condon factor) for this hopping with the phonon coordinate
in its quantum mechanical ground state (see Eq.~\ref{e18}). For
large exciton-phonon coupling, we have
$\exp(-\chi^2/(\hbar\omega)^2)\rightarrow 0$, and Eq.~(\ref{e22a})
would equate Eq.~(\ref{e13a}). In that limit, the polaron
dispersion relation would become flat (i.e. the polaron energy
would be independent on wavevector $k$), and truly spatially
localized solutions, as we obtain them from the DNLS, could indeed
be eigensolutions of the quantum Hamiltonian. This is the
classical limit, where tunneling is not possible and the hopping
probability becomes zero! However, for small exciton-phonon
coupling, the polaron energy is lowered by an additional term $2J
\exp(-\chi^2/(\hbar\omega)^2)$, which may render the polaron more
stable than the exciton, even when the DNLS would predict the
opposite.

It is instructive to compare the results of this chapter with
those of the coherent state treatment. We have seen in
Sec.~\ref{sub3.2} that the product ansatz Eq.~\ref{e24.8} and
~\ref{e24.9} directly leads to the semiclassical DNLS without
invoking any further approximation. The DNLS, however, fails in
certain parameter regimes where $\chi/\hbar\omega<\approx1$. The
essential difference between coherent state treatment and the
approach discussed above is that the latter uses a much larger
basis (Eqs.~\ref{eqbasis1} and ~\ref{eqbasis2}). In particular,
the 'basis' of the coherent state treatment is local in the sense
that the polaron is confined to the exciton by construct. In
contrast, the basis Eqs.~\ref{eqbasis1} and ~\ref{eqbasis2} allows
phonon excitations away from the exciton with, e.g. a phonon
sitting at site 1, while the exciton is sitting at site 0 at the
same moment (i.e. $n_{+1}=1$). Counterintuitively, that phonon at
site 1 does \textit{not} interact with the part of the exciton
that is also residing on site 1 when shifting $\Phi$
(Eq.~\ref{eqbasis3}) by one site along the lines of the Bloch
ansatz Eq.~\ref{eqbasis1}. There is no classical counterpart to
this phenomenon; it reflects the non-local character of quantum
mechanics. In the limit $\chi/\hbar\omega\gg1$, the nonlocal
character become less important, and the DNLS becomes a good
approximation.

\section{Pump-probe experiments on ACN}
\subsection{Experimental Results}
\label{sec:p1}

\begin{figure}[t]
\centerline{\epsfig{figure=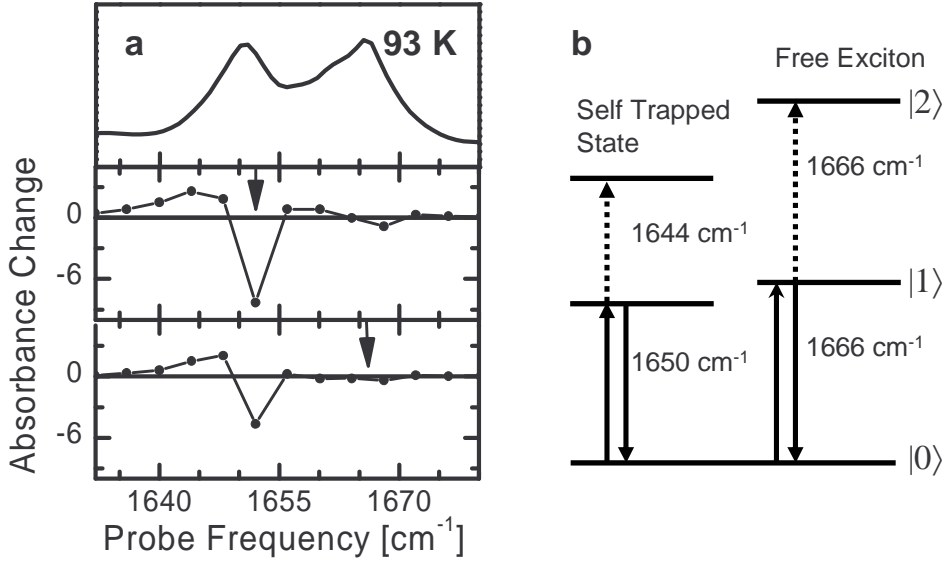,
      width=.9\columnwidth, bbllx=50 pt,
  bblly=420 pt,bburx=560 pt,bbury=743 pt, clip=}}
\caption{(a) Linear absorption spectrum (top) and pump probe
spectra of the C=O mode of crystalline ACN at 93 K for two
different narrow band pump pulses chosen to be resonant with each
of the absorption bands. The arrows mark the center frequency of
the pump pulse, which was spectrally narrow enough to pump only
either one of the transitions. (b) Level scheme of the system,
explaining the distinctively different response of both modes.
Adapted from Ref.~\cite{edl02b}.}\label{figp3}
\end{figure}

With these introductory words in mind, we now turn to our recent
femtosecond pump-probe experiments on the C=O band of crystalline
ACN~\cite{edl02b,edl03}. (The pump-probe response of the NH band
of ACN has been studies as well~\cite{edl02a,edl04}, which however
shall not be discussed here.) In such a pump-probe experiment, the
various modes are first excited with an intense short IR laser
pump pulse, and the response upon that excitation is then probed
by a second short, but much weaker probe-pulse. The information
gain of femtosecond pump-probe experiments, as compared to simple
IR absorption spectroscopy (Fig.~\ref{fig2}), is manifold:
\begin{itemize}
\item The anharmonicities of the individual modes can be measured,
as sketched in Fig.~\ref{figp3}b: When pumping a state, its first
excited state is populated, and the probe pulse is then probing
both the 1--2 upward transition and the 1--0 downward transition.
If the state is harmonic, both upward and downward transition
appear at exactly the same frequency, but with opposite signs, and
cancel each other exactly. In contrast, if the state is
anharmonic, they lead to two bands with alternating signs
separated by the anharmonicity of the transition.

\item If two states are coupled, excitation of the one state will
lead to an response (frequency shift) of the other. It shows to
what extend the various transitions are entangled, or whether they
can be viewed as separated problems. If, for example, the two
lines in ACN would originate from molecules in different
conformations or surroundings (topological defects), as has been
suggested various times~\cite{bla85,fan90}, pumping of the one
state would not cause any response of the other state (in contrast
to what is seen in the experiment). On the other hand, in the case
of a Fermi-resonance, another often discussed alternative
explanation of the doublet in the ACN spectrum~\cite{joh85}, a
very characteristic coupling pattern is expected. Investigating
such coupling patterns in detail, compelling evidence against both
of these possibilities could be provided~\cite{edl03}.

\item The lifetime of the pumped states can be measured with
femtosecond time resolution. In this way, the original idea of
Davydov~\cite{dav77} can be tested, namely to what extent
vibrational self-trapping may stabilize the excitation for
timescales that would allow to make the energy available for
subsequent biological processes.
\end{itemize}

Fig.~\ref{figp3}a shows the result of pump-probe experiments on
the two bands of crystalline ACN~\cite{edl02b}: When resonantly
pumping the "anomalous" band (1650~cm$^{-1}$) of ACN, the band
bleaches (negative response) and a positive band emerges at
1644~cm$^{-1}$. In contrast, when resonantly pumping the "normal"
band (1666~cm$^{-1}$), hardly any bleach of the band itself is
observed. Nevertheless, the anomalous band does responds with a
signal which is similar in shape, but is slightly smaller than
when pumping it directly.

In a somewhat heuristic approach, Edler et al. have discussed the
distinctly different responses of the two bands of ACN in
Ref.~\cite{edl02b}. The self-localized state was treated along the
lines of Scott's theory~\cite{sco85b}, which would relate the
anharmonicity of the anomalous band to the $-\chi^2/\hbar \omega
{B_j}^\dag{B_j}^\dag B_j B_j$ term in Eq.~(\ref{e19}). At the same
time, it was assumed that one can 'cut out' the exciton part of
the Hamiltonian $H_{ex}$ (Eq.~\ref{e2}) from the remainder by
writing:
\begin{eqnarray}
H_{ex}=\hbar \Omega \sum_{j=1}^N \left({ B_J}^\dag B_j + 1/2
\right) - J \sum_{j=1}^N \left({ B_j}^\dag B_{j+1} + {B_j}^\dag
B_{j-1} \right)
\end{eqnarray}
The discussion of the exciton is a little more involved: It turns
out that, apart from the artificial separation from the
phonon/polaron part, this Hamiltonian is not a very realistic
description of molecules. A chemical bond is not harmonic in
reality, but is better described by a Morse-potential. Vibrations
of real-worlds molecules are intrinsically anharmonic (on-site
anharmonicity). For example, the anharmonicity of a C=O vibrator,
defined as the difference in frequency of the 0-1 and the 1-2
transition, lies between 10-20~cm$^{-1}$. Hence, we should use
instead for the exciton Hamiltonian:
\begin{eqnarray}
H_{ex}=\hbar \Omega \sum_{j=1}^N \left({ B_J}^\dag B_j + 1/2
\right) - \Delta \cdot {B_j}^\dag{B_j}^\dag B_j B_j-
 J \sum_{j=1}^N \left({ B_j}^\dag B_{j+1} + {B_j}^\dag B_{j-1} \right)\label{ep2a}
\end{eqnarray}
where $\Delta$ is the on-site anharmonicity. Note that the
additional term does not affect the one-exciton states nor any
polaron solution in the one-exciton manifold where
${B_j}^\dag{B_j}^\dag B_j B_j=0$, which is why we normally can
ignore it. It also doesn't affect the outcome of normal absorption
spectroscopy along the lines of Fig.~\ref{fig2}. The term however
does affect the way how one potentially sees one-exciton states in
pump-probe spectroscopy, since the two-exciton manifold is reached
in such experiments.

In light of this discussion, the experimental result shown in
Fig.~\ref{figp3}a, bottom, is in fact surprising. In this
experiment, a vibrational state (the higher frequency band) is
pumped, but cannot be depleted. Similar pump-probe experiments
have been done by Hamm and coworkers on hundreds of molecules, but
a result of this sort has never been obtained. 'Normal'
vibrational states are intrinsically anharmonic, and as a result,
one  expect to see a pump-probe response similar to that of the
'anomalous' band, i.e. a bleach/stimulated emission at its
original 0--1 position and a positive 1--2 absorption frequency
shifted by the intrinsic anharmonicity $\Delta$. In
Refs.~\cite{edl02b,edl03}, several alternative explanations have
been excluded, and the only possible explanation left over was:
The higher frequency peak in the absorption spectrum of ACN
behaves as if it were harmonic, despite the fact that it is
composed of C=O vibrators that are intrinsically anharmonic. Only
if the state is effectively harmonic, upward and downward
transitions in Fig.~\ref{figp3}b could cancel exactly and the
zero-response seen in the experiment is expected.

Interestingly, one indeed finds that an exciton may behave
effectively harmonic even in the presence of on-site
anharmonicity, provided its delocalization length is sufficiently
long. In fact, one can show by a perturbative expansion of the
Hamiltonian Eq.~(\ref{ep2a}),  treating the anharmonicity $\Delta$
as small, that the effective anharmonicity of an exciton scales
like the so-called participation ratio~\cite{edl02b}:
\begin{eqnarray}
\Delta_{eff}=\Delta\sum_j\Psi_j^4
\end{eqnarray}
where $\Psi_j$ are the expansion coefficients of the exciton in a
site basis. The participation ratio is a commonly used measure of
the degree of delocalization, i.e. $\sum_j\Psi_j^4=1$ for a fully
localized state and $\sum_j\Psi_j^4\rightarrow0$ for a perfectly
delocalized state. The exciton's anharmonicity is a direct measure
of its degree of delocalization which, according to
Fig.~\ref{figp3}a, bottom, is large at 93~K.

\begin{figure}[t]
\centerline{\epsfig{figure=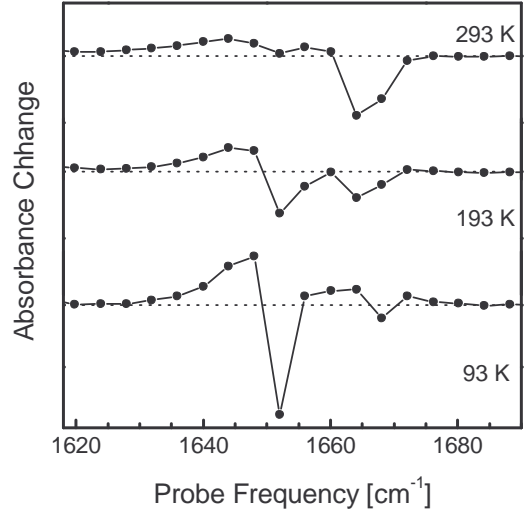,
      width=.5\columnwidth, bbllx=70 pt,
  bblly=520 pt,bburx=290 pt,bbury=740 pt, clip=}}
\caption{Pump-probe spectra of the C=O mode of crystalline ACN as
a function of temperature from 93 K to room temperature. Adapted
from Ref.~\cite{edl02b}.}\label{figp4}
\end{figure}

Interesting in this context is the temperature dependence of this
effect (Fig.~\ref{figp4}): At low temperatures, in the absence of
disorder, the exciton is fully delocalized, and one can hardly
bleach it. However, as temperature rises, disorder increases and
the exciton starts to Anderson-localize. As a result, the exciton
transition could then indeed be bleached because its effective
anharmonicity increases (Fig.~\ref{figp4}, top). The onset of
disorder-induced Anderson localization of the free exciton is
observed at about the same temperature where the intensity of the
self-trapped state disappears. This finding suggests that both
localization mechanisms are connected. Disorder localization is
caused by random variations of the diagonal elements of the
excitonic coupling Hamiltonian Eq.~(\ref{ep2a}). The amide I
frequency is known to vary significantly with hydrogen bond
distance (on the order of 20-30 cm$^{-1}$/\AA); hence, thermal
excitation of phonons that give rise to random fluctuations of the
hydrogen bond distances lead to disorder-induced localization of
the free exciton. Since these are exactly the phonons which also
mediate self-trapping at low temperatures, thermal excitation at
the same time diminish the intensity of the self-localized state.
Hence, it is the same nonlinear interaction, the variation of the
amide I excitation energy with hydrogen bond distance, which gives
rise to the two effects: (i) diagonal disorder and as a
consequence, disorder-induced localization of the free exciton
with rising temperature and (ii) self-localization at sufficiently
low temperatures. The competition between Anderson (disorder)
localization and nonlinear self-localization in the presence of
disorder has been investigated in Ref.~\cite{feddersen91}: Those
states that are self-localized for large enough nonlinear
exciton-phonon coupling $\chi$ are quite distinct from those that
are Anderson (disorder) localized. Upon reducing the nonlinear
coupling, the self-trapped states become less localized and
eventually reach bifurcation points, below which they do not
exist. Surprisingly, the states that are Anderson localized for
vanishing nonlinear coupling become rapidly delocalized and
unstable as nonlinear coupling is increased just slightly.

Hamm and coworkers~\cite{edl02b} concluded that the pump-probe
experiments are fully consistent with earlier interpretations of
conventional temperature dependent absorption
spectroscopy~\cite{sco92}, namely that the higher frequency band
of ACN relates to an free exciton whereas the lower frequency band
relates to a self-localized state. Nevertheless, it turns out that
the lifetime of either of these sates is short, 2
ps~\cite{edl02b}, and as such, the original idea of
Davydov~\cite{dav77} that vibrational self-trapping stabilizes the
excitation, did probably \textit{not} come true. A lifetime of 2
ps is a typical value for vibrational energy dissipation in
condensed phase systems, and vibrational self-trapping is not
capable to extend this lifetime by any significant amount. 2 ps
appears to be too short to trigger any serious conformational
transition of a protein that could potentially be important in the
biological function of the system.

\subsection{Comparison to a Numerically Exact Theory}

Although the discussion outlined above seemed to explain the
experiment conclusively, we note again that the 'theory' of
Ref.~\cite{edl02b} artificially tried to 'cut out' the exciton
part from the remainder. At the same time, it is clear that the
free exciton is not an eigenstate of the full Hamiltonian
Eq.~(\ref{e1}--\ref{e4}), which can be seen from simple insertion.
Furthermore, also the experiments show that exciton and polaron
are, of course, coupled to each other, leading to a spectral
response of the latter when pumping the former (Fig.~\ref{figp3}a,
bottom). Unfortunately, none of the limiting regimes discussed in
the previous sections apply to the parameter range that has been
proposed for the ACN problem ($\chi\approx25$~cm$^{-1}$,
$\omega\approx$50~cm$^{-1}$, $J=$4-10~cm$^{-1}$):
\begin{itemize}
\item The DNLS (Sec.~\ref{s3} and ~\ref{sec:2.1}) is valid in the
limit $\chi/\hbar\omega\rightarrow\infty$, in which case the
polaron binding energy $E_{b}=-\chi^2/\hbar\omega$ would become
large compared to the discreteness of the oscillation frequency
$\hbar\omega$. However, for the given set of parameters for ACN we
find $\chi/\hbar\omega=0.5$; by no means large! In fact,
Fig.~\ref{figp2} shows that polaron solutions are expected in this
regime for relatively large exciton couplings $J$, where the DNLS
would strictly predict exciton solutions.

\item The small $J$ limit seems to be better justified
(Sec.~\ref{sec:1.1} and ~\ref{sec:1.2}). Yet, this limit does not
predict the existence of a free exciton, in disagreement with the
common picture of the spectroscopy of ACN, and in disagreement
with the interpretation of the experimental pump-probe results
discussed above. According to the small $J$ limit, one would
rather expect a first excited state that lies exactly one phonon
quantum above the ground state (Eq.~\ref{e13a}). However, a
hypothetical phonon frequency of $\omega=15$~cm$^{-1}$ (the
spacing between two bands in Fig.~\ref{fig2}) would result in a
completely different temperature dependence with a much lower
turn-over between low and high temperature regime~\cite{sco92},
and would furthermore contradict the experimental observation that
it is indeed a $\omega=50$~cm$^{-1}$ phonon that couples strongly
to the hydrogen bond~\cite{edl02a,edl04}.
\end{itemize}

The existence of the two solutions of the DNLS, an extended free
exciton and an self-localized small polaron, was commonly used to
rationalize the two peaks in the absorption spectrum of ACN
(Fig.~\ref{fig2}a). According to this interpretation, the lower
frequency band relates to a self-trapped state with a temperature
dependence given by the displaces oscillator
picture~\cite{ale85,ale86,sco92} while the higher frequency band
relates to the free exciton. Again, the free exciton is
\textit{not} an eigenstate of the full-quantum Hamiltonian
Eq.~(\ref{e1}--\ref{e4}). Hence, we are in the bizarre situation
that the DNLS apparently describes the experimental results
correctly, knowing that we are in a parameter regime where it is
\textit{not} valid, while the formally more accurate full-quantum
Hamiltonian does not. In light of this discussion, it might seem
that the correct result of the DNLS is an \textit{unfortunate
coincidence}.

\begin{figure}[t]
\centerline{\epsfig{figure=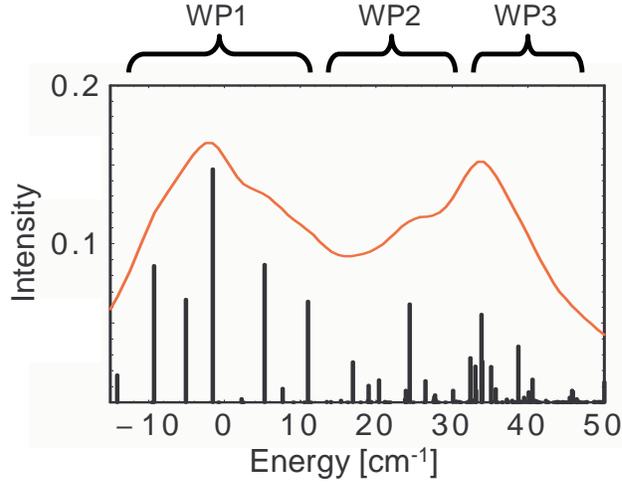,
      width=.6\columnwidth, bbllx=90 pt,
  bblly=420 pt,bburx=500 pt,bbury=750 pt, clip=}}
\caption{Absorption spectrum of the Holstein in 3D with
$\hbar\omega=50~\rm{cm}^{-1}$, $\chi=-25~\rm{cm}^{-1}$,
$J_\parallel=10~\rm{cm}^{-1}$ and $J_\perp=-10~\rm{cm}^{-1}$.
Stick spectra and spectra convoluted with a Lorentzian lineshape
function with width $10~\rm{cm}^{-1}$ (FWHM) are shown. The
eigenstates appear in clusters with properties of a polaron in its
ground state (WP1), an exciton in its ground states (WP2), and a
polaron in its first excited state (WP3, see Fig~\ref{figp6}).
Adapted from Ref.~\cite{ham06a}.}\label{figp5}
\end{figure}

In order to ultimately resolve this issue, Hamm et al. set out to
calculate the absorption spectrum of ACN on a numerically exact
level~\cite{ham06a}, along the lines of Sec.~\ref{sec:p2}. In
doing so they were guided by the DNLS result that polaron and free
exciton may coexist only in 3D, in which case barriers separate
both solutions (see Sec.~\ref{sec:p2}), and hence render an
exciton solution meta-stable, even if it is slightly higher in
energy. This is in accordance with the well known result from a
simple quantum mechanical treatment of a square well potential in
three dimensions, where a bound state appears only if the product
$E_{0}a^2$ exceeds a threshold value (with $E_{0}$ the depth and
$a$ the width of the well). In contrast, an one-dimensional square
well has at least one bound state for any value of
$E_{0}a^2$~\cite{schiff59}.

Hamm et al. were furthermore guided by the observation obtained
from inspection of the 3D structure of the ACN crystal
(Fig.~\ref{fig1}a) that exciton coupling along the hydrogen bond
direction is negative ($J_\parallel>0$), while the coupling
between hydrogen bond chains is expected to be positive ($J_\perp
<0$). This is since exciton coupling is dominated by through-space
transition-dipole coupling~\cite{kri86,tor92}:
\begin{eqnarray}
J=-\frac{\vec{\mu_i}\cdot
\vec{\mu_j}}{r_{ij}^3}+3\frac{(\vec{\mu_i}\cdot
\vec{r_{ij}})(\vec{\mu_j}\cdot \vec{r_{ij}})}{r_{ij}^5},
\label{eqdipoledipole}
\end{eqnarray}
where $\vec{\mu_{i}}$ and $\vec{\mu_{j}}$ are the C=O transition
dipoles of two sites (as vectors) and $\vec{r_{ij}}$ is a vector
connecting them. The nearest neighbor distance $r_{ij}$ along the
hydrogen-bonded chain is essentially the same as that between
chains~\cite{brown54}, which is why inter- and intra-chain
couplings are expected to be of the same order, albeit with
opposite signs. There is no reason to assume that inter-chain
coupling can be neglected.

Introducing anisotropy into the exciton coupling increases the
parameter space considerably. In Ref.~\cite{ham06a}, an extensive
search in parameter space was performed, staying within limits
that seemed physically meaningful for the ACN problem, and looking
for full-quantum solutions that predict coexistence of free
exciton and polaron. Fig.~\ref{figp5} shows the best result that
could be obtained. The total oscillator strength is distributed
over many eigenstates, nevertheless, the oscillator strength
clusters in a way that two dominant bands remain when convoluting
the stick spectrum with a Lorentzian lineshape function of width
$10~\rm{cm}^{-1}$ (FWHM). The thus obtained spectrum resembles the
experimental 30~K spectrum reasonably well (Fig.~\ref{fig2}). The
separation of the two bands ($\approx20-30~\rm{cm}^{-1}$) is
somewhat larger than in the experiment, yet it clearly is smaller
than one phonon quantum $\hbar\omega$. Furthermore, the higher
frequency band shows some substructure in the simulation, similar
to the experiment.

\begin{figure}[t]
\centerline{\epsfig{figure=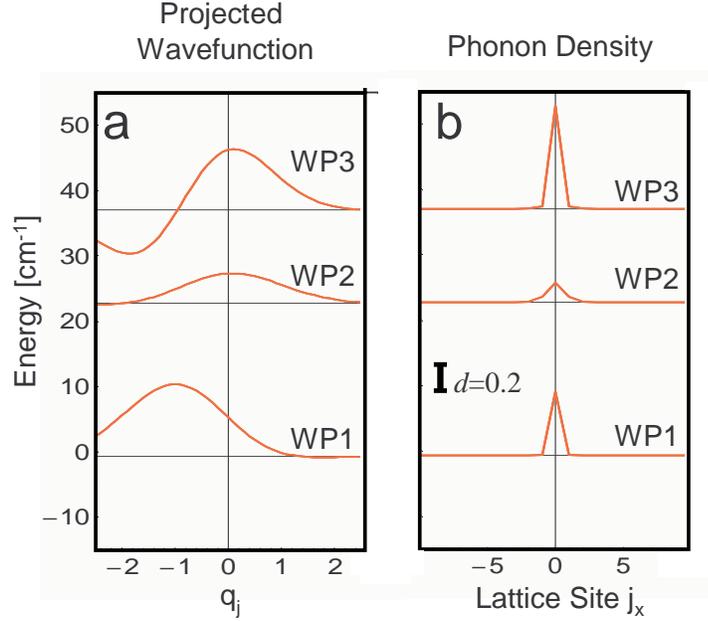,
      width=.9\columnwidth, bbllx=60 pt,
  bblly=400 pt,bburx=530 pt,bbury=740 pt, clip=}}
\caption{(a) Projection of (t=0) wavepackets onto phonon
coordinate $q$ and (b) the corresponding phonon-densities
$d(j)=\langle\Psi|B_0^{\dag} B_{0}b_j^{\dag} b_{j}|\Psi\rangle$.
Wavepackets WP1, WP2 and WP3 are constructed from the embraced
eigenstates as shown in Fig.~\ref{figp5}. The parameters in this
simulation were: $\hbar\omega=50~\rm{cm}^{-1}$,
$\chi=25~\rm{cm}^{-1}$, $J_\parallel=10~\rm{cm}^{-1}$,
$J_\perp=-10~\rm{cm}^{-1}$. Adapted from
Ref.~\cite{ham06a}.}\label{figp6}
\end{figure}

We pause and note that we do not observe stationary eigenstates in
a femtosecond pump-probe experiment (which are not resolved anyway
in reality due to broadening mechanisms). Rather, a femtosecond
laser pulse is spectrally broad (due to its shortness) and
impulsively excites a coherent superposition state (wavepacket) of
all eigenstates that are covered by the spectral width of the
pump-pulse. The typical width of the pulses used in
Refs.~\cite{edl02b,edl03} was 10~cm$^{-1}$, which is why it makes
sense to discuss the properties of the wavepackets WP1, WP2 and
WP3 (as indicated in Fig.~\ref{figp5}), rather than that of
individual eigenstates.

A closer analysis shows that the eigenstates that cluster in the
lower frequency band (WP1) resemble a ground state harmonic
oscillator wavefunction with an origin displacement of about 1
(see  Fig.~\ref{figp6}a), which is somewhat larger than the value
predicted from the simple displaced oscillator picture with
$q=\sqrt{2}\chi/\omega\approx0.71$. The eigenstates that cluster
in the higher frequency band, on the other hand, can be classified
in the following way: The lower energy portion of these states,
clustering as a shoulder in the absorption spectrum (WP2),
resembles ground state harmonic oscillator wavefunctions
\textit{without any displacement}, while the higher energy portion
(WP3) resembles first excited state harmonic oscillator
wavefunctions, again \textit{with} origin shift.

Fig.~\ref{figp6}b shows the density of phonons around an exciton,
calculated as
\begin{eqnarray}
d(j)=\langle\Psi|B_0^{\dag} B_{0}b_j^{\dag} b_{j}|\Psi\rangle
\end{eqnarray}
which reveals essentially the same result: Wavepackets WP1 and WP3
represent phonons dressed around an excitons, i.e. ground and
first excited state of a polaron, whereas wavepacket WP2 has much
smaller phonon contribution (albeit not zero). Hence, the results
seem to be in reasonable agreement with experimental observation.
However, it should be noted that such free-exciton like solutions
are found only in a very limited parameter regime. In particular,
they are found only when exciton couplings with alternating signs
of $J_\parallel$ and $J_\perp$ are introduced. The appearance of
the double peak structure in the absorption spectrum of
crystalline ACN thus seems to depend relatively critically on a
balance of parameters, which might explain why vibrational
self-trapping is observed so clearly only in ACN, while related
crystals show, if at all, only minor effects~\cite{edl04}. It is
important to note that equivalent exciton-like solutions could not
be found in 1D, where the displaced oscillator picture
(Sec.~\ref{sec:1.2}) remains intact even for relatively large
exciton couplings $J$.

\begin{figure}[t]
\centerline{\epsfig{figure=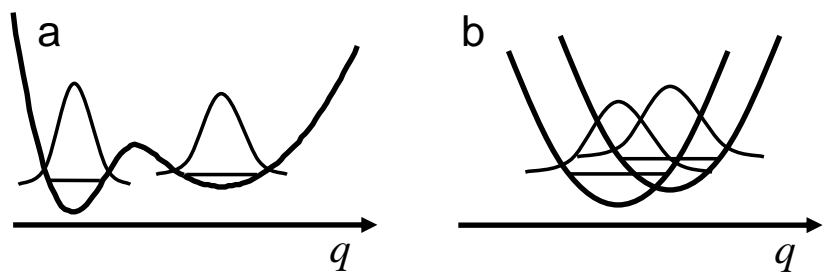,
      width=.8\columnwidth, bbllx=40 pt,
  bblly=490 pt,bburx=530 pt,bbury=670 pt, clip=}}
\caption{Two mechanisms that might lead to coexistence of exciton
and polaron solutions: (a) two regimes on one effective
(adiabatic) potential energy surface separated by a barrier which,
however, requires that the length scale of quantum mechanical
delocalization of the wavefunction is smaller than that of the
barrier, or (b) two effective (adiabatic) surfaces.}\label{figp7}
\end{figure}

In Sec.~\ref{sec:p2} we have seen that exciton and polaron
solutions may coexist in certain parameter regimes in the 3D case,
as a result of a barrier that separates the two regions
(Fig.~\ref{figp1}). We find such a barrier only in the regime
where the adiabatic DNLS becomes valid, i.e. for $\chi/\hbar
\omega\gg 1$, while the transition between exciton and polaron is
smooth in the quantum regime (Fig.~\ref{figp2}). This can be
rationalized as follows: In the quantum case ($\chi/\hbar
\omega<1$), the potential phonon displacement $q$ is of the order
of one or smaller, where the phonon displacement is defined in
units of quantum mechanical delocalization length. In other words,
quantum-mechanics would average out any barrier that occurs on
length scales smaller than the delocalization length of the phonon
wavefunction. In that sense, the coexistence of polaron and
exciton solutions found in the numerically exact solution of the
Holstein Hamiltonian (Fig.~\ref{figp6}) is due to an effect that
is different from the one discussed in Sec.~\ref{sec:p2}: It is
not a barrier on an effective (adiabatic) ground state surface
(Fig.~\ref{figp7}a), but must be related to at least two adiabatic
surfaces, as sketched in Fig.~\ref{figp7}b. A physical picture of
these two surfaces, apart from the black-box result from the
numerically exact diagonalization of the full Hamiltonian, still
remains to be developed. Nevertheless, clearly it is the 3D nature
of the problem that is responsible for exciton-like solutions. In
1D, the displaced oscillator picture (Sec.~\ref{sec:1.2}), which
does not predict any exciton solutions, remains intact in a wide
parameter range.

\section{Summary and Conclusions}
From its first appearance in the literature in 1983, the
temperature dependent peak of the C=O vibrations in acetanilide
has been associated to the Davydov soliton idea.  The latter assumes
that when excitons are coupled strongly to acoustic phonons a
soliton-like excitation forms that may propagate ballistically.
Although Scott and collaborators used optical phonons for
acetanilide, their selftrapping approach was similar to that of Davydov.
The Davydov soliton idea has been very appealing from the very beginning
since it asserts that
nonlinear type modes play an important role in energy transfer in
biology.  Nonlinearity, of course, is not unknown to biology; on the contrary
most functional motions of macromolecules are large amplitude ones
and thus nonlinear.   The theoretical as well as experimental
fascination with acetanilide over a period of more than two
decades results fundamentally from its similarity to peptide
systems.  If soliton-like objects  exist in acetanilide then one
may investigate their presence and function in  proteins.

On the theoretical side, the basic model of Scott for acetanilide
involves the following elements: (i) The system is assumed to be
one dimensional.  This is because the C=O dipoles along the
hydrogen-bond axis seem to interact stronger among them than with
the ones in the perpendicular directions. (ii)  A bear vibrational
C=O excitation that tunnels slowly from bond to bond ($J=
5~\rm{cm}^{-1}$) couples to a single optical phonon  (or to few
phonons) of frequency $\omega = 50~\rm{cm}^{-1}$  with coupling
$\chi = 25~\rm{cm}^{-1}$.  As a result a localized  small polaron
of the type of a DNLS discrete breather forms. (iii) The double
absorption peak in the amide I region results from the polaron
state at $1650~\rm{cm}^{-1}$ and the free exciton at
$1665~\rm{cm}^{-1}$ (iv) The temperature dependence of the polaron
state arises from the Frank-Condon factors connecting the
unexcited phonon ground state with the displaced phonons in the
excited polaron state. This theoretical Scott picture that is
adiabatic in nature since it employs the Davydov methodology, was
corroborated by Alexander and Krumhansl with the modification that
acoustic phonons are also involved.

In order to converge to the polaron picture, numerous alternative
explanations had to be eliminated.  The most serious objection to
the polaron comes for the Fermi-resonance idea, viz. the
accidental degeneracy of  the amide I symmetric stretch mode with
a combination band of an in-plane N-H bend deformation as well as
a torsional mode~\cite{joh85}.  Although Raman as well as neutron
scattering experiments by Barthes and collaborators largely
eliminated this possibility, the mere fact that upon complete
deuteration the anomalous peak disappears, presents a yet
unexplained feature that could speak in favor of a Fermi
resonance~\cite{bar91,bar92}. An alternative issue is related to
the active presence of acoustic modes in selftrapping; although
there is indication of their absence from incoherent neutron
scattering, there is no direct measurement yet of the presumed
optical modes that are excited in the processes of the C=O
vibrations.   While these two experimental issues seem to be still
pending all other questions regarding  the role of  low IR modes,
presence of defects or possible hydrogen bond non-degeneracies
have been conclusively addressed by the group of Barthes.

While most of the absorption  and scattering experiments were done
in the eighties and nineties, the recent pump-probe experiments of
Hamm and collaborators provide more direct information on the
nature of the states involved.  Through experiments in the
acetanilide N-H band was possible to identify a $50~\rm{cm}^{-1}$
hydrogen bond related  mode that assists selftrapping in this
band.  One then makes the natural assumption (compatible with the
earlier work) that this phonon is responsible for the amide I
selftrapping as well. Furthermore, in pump-probe experiments in
the amide I region itself, they found that while the low frequency
$1650~\rm{cm}^{-1}$ band corresponds indeed  to a nonlinear mode,
the $1665~\rm{cm}^{-1}$ one is certainly a linear one. As result,
the former may be assigned to a selftrapped polaron state while
the latter to a free exciton one, in a manner compatible with
adiabatic theory of Scott.

This adiabatic theoretical polaron scheme has been challenged  recently
by an exact diagonalization performed on the Holstein
Hamiltonian as applied to the specific acetanilide problem.  Using
a numerically exact procedure Hamm et al.  found
that while the polaron state could be identified as the lowest
energy state, the free exciton state did not appear in the
calculated spectrum~\cite{ham06a}.
These numerics showed that there is no free exciton state neither
in one nor in three dimensions for  the "standard" acetanilide
parameter values with hopping rate in the range
$J \approx 5-10~\rm{cm}^{-1}$.
In order to bypass this problem, Hamm et al.  introduced
different signs in the hopping rares depending on direction, viz.
negative overall rates along the hydrogen-bonded axis while
positive along the plane perpendicular to it.  The resulting
spectrum is then compatible with the experiment within the standard
parameter regime.

The issue of the free exciton state has been discussed extensively
in the present work.
Using the semiclassical
theory of Kalosakas et al., we showed that in three
dimensions free exciton and polaron states are separated by a
barrier while there is no such barrier in one dimension~\cite{kal98a}.
Furthermore, for the standard acetanilide parameter regime the one
dimensional semiclassical free exciton state has quite a short
lifetime and is in general unstable.  Thus, even semiclassically, the
one-dimensional acetanilide picture needs to be revised.  Using the
double well adiabatic potential for free exciton and polaron and upon an
ad-hoc requantization we find that the resulting two lowest states have
polaron and free-exciton features respectively.  Thus, within this
intuitive, yet approximate, 3D picture we recapture in principle
Scott's original explanation.  The problem is that this result does
not appear to agree quantitatively with the exact numerics mentioned
above, unless different sign hopping terms are used in the latter.

Cristaline acetanilide  proved to be a good medium for experiment
as well as theory for over twenty five years.  During this period
it permit the application of diverse experimental techniques as
well as the testing of a large variety of theoretical ideas.  It
is clear that although many issues have been addressed and
resolved, there are still many challenges ahead. In addition to
the central dimensionality issue related to the free exciton, the
absence of a temperature dependent peak from the fully deuterated
acetanilide or the dispersion features of the modes that assist
selftrapping need to be understood better.  Furthermore,
selftrapping in the N-H bond needs also  to be addressed that is
actually different compared to that in the C=O bonds.
Specifically, while a phonon progression that is absent in the
later is seen in the former, it nevertheless terminates abruptly
in a way dissimilar to what is expected from Eq.~\ref{e13a}. And
while the C=O overtones where treated by Scott through QDNLS, it
would be interesting to analyze them directly through the Holstein
model. It is clear that the exciting "acetanilide story" is far
from over.

{\bf Acknowledgment} This work is dedicated to the memory of Al
Scott who has been a source of inspiration  for us in this and may
other problems of nonlinear physics. P.H. acknowledges financial
support from the Deutsche Forschungsgemeinschaft (collaborative
research center SFB 450) as well as from the Swiss Science
Foundation (grants 2100-067573/1 and 200020-107492/1) and GPT
acknowledges grant 2006PIV10007 of the Generalitat de Catalunia,
Spain. We wish to thank Julian Edler for important contributions
to this work.

\section{APPENDIX : Brief historical guide to the research in ACN}
\label{sec:ap2}
Due to the numerous number of works published on the "Davydov soliton" as well as ACN, it is often hard to
keep track of the various contributions and the significant developments in the "acetanilide story".  We present here a partial list of contributions in a historical order that are specifically relevant  to the acetanilide developments.
\begin{itemize}
\item{1973} Davydov introduces the idea that as a result to
coupling with acoustic phonons a vibrational excitation may
selftrap and become a non-topological soliton~\cite{dav73}. Careri
finds experimentally an "anomalous peak" in
acetanilide~\cite{car73}. \item{1982} Davydov publishes his book
on "Biology and Quantum Mechanics" where he details his approach
to energy transfer in biomolecules~\cite{dav82}. \item{1983}
Careri, Scott and collaborators present experimental and
theoretical analysis of the C=O band of crystalline
acetanilide~\cite{car83}. After eliminating possible alternative
explanations they conclude  that the temperature dependent ACN
peak at $1650~\rm{cm}^{-1}$ is due to a "Davydov soliton" type
mechanism. \item{1984} Careri, Scott and collaborators present
detailed experimental and theoretical details on the acetanilide
"anomalous peak"~\cite{car84}. Eilbeck, Lomdahl and Scott analyze
selftrapping of the amide I mode through the Discrete Selftrapping
Equation~\cite{eil84}. \item{1985} Alexander treats theoretically
the Davydov Hamiltonian using Holstein polaron
techniques~\cite{ale85}. Blanchet and Fincher find experimentally
that low IR as well as a band at   $3250~\rm{cm}^{-1}$  have  the
same temperature dependence as that of the "anomalous peak" and
suggest that a topological defect might be responsible for the
anomalous amide structure~\cite{bla85}. Johnston and Swanson
propose that that the anomalous peak is a result of Fermi
resonance between the amide I mode and a combination mode between
an N-H motion and a strongly temperature dependent low frequency
phonon~\cite{joh85}.  Scott and collaborators introduce the
quantum DST equation and explain higher frequency anomalous lines
as overtone spectra of the C=O band~\cite{sco85b}. \item{1986}
Takeno introduces a classical vibron model and studies the
acetanilide temperature dependence in a way similar to that of
localized modes in alkali halides~\cite{tak86}. Alexander and
Krumhansl provide an adiabatic treatment of the Davydov
Hamiltonian and analyze the ACN peak using an acoustic phonon
chain~\cite{ale86}. \item{1988} Barthes and coworkers study
experimentally acetanilide and while confirm the presence of
anharmonicity in the solid they report no evidence of acoustic
phonons involved in selftrapping~\cite{bar88}. \item{1989} Scott
uses the theory of color centers and proposes a quantitative
explanation for the temperature dependence of the "anomalous" ACN
peak~\cite{sco89}. \item{1990} Austin and collaborators perform
pump-probe experiments in acetanilide and propose that the
temperature  dependence of the anomalous peak is due to hydrogen
bond non-degeneracy~\cite{fan90}. \item{1991} Barthes and
collaborators show through inelastic neutron scattering
experiments that ACN data are not consistent with a topological
defect assumption~\cite{bar91}. They also claim that no indication
of acoustic phonons are seen in the low IR spectra. Furthermore
through Raman scattering experiments  they report that while fully
deuterated acetanilide shows no anomalous peak, a partially
deuterated one does have one, downshifted however by
about$150~\rm{cm}^{-1}$ from the main amide I peak. \item{1992}
Scott summarizes the "Davydov soliton" and the "acetanilide story"
in a review article in Physics Reports~\cite{sco92}. \item{1995}
Barthes and collaborators use neutron scattering experiments and
find no non-degeneracy in the hydrogen-bonded
proton~\cite{john95}. \item{1999} Kalosakas, Aubry and Tsironis
investigate the semiclassical Holstein model and propose that
phononic normal modes of the semiclassical polaron may be related
to low frequency ACN modes~\cite{kal98a,kal98b}. \item{2000}
Barthes and collaborators show that the low IR modes investigated
earlier have no anomalous temperature dependence~\cite{spi00}.
\item{2002} Hamm, Edler and Scott use pump-probe experiments and
analyze the NH spectrum of ACN~\cite{edl02a}. They identify a
phonon of approximately $50~\rm{cm}^{-1}$ that seems to be
involved in selftrapping of the exciton. Hamm and Edler  publish
pump-probe experiments for the C=O band that identify the
"anomalous" peak as a nonlinear state and the "normal" one as a
linear free exciton state~\cite{edl02b}. \item{2006} Hamm and
collaborator present a numerical diagonalization of the Holstein
model in one and three dimensions and find inconsistencies between
numerical results in one dimension and standard adiabatic
anomalous peak explanation. They propose that the C=O polaron are
a result of the three dimensional nature of the ACN crystal as
well as specific exciton dipole-dipole matrix
elements~\cite{ham06a}.

\end{itemize}


\end{document}